%
%
%
%

\catcode `\@=11 

\def\@version{1.4}
\def\@verdate{22nd Feb 1994}

%
%
%
%


\newif\ifprod@font

\ifx\@typeface\undefined
  \def\@typeface{Comp. Modern}\prod@fontfalse
\else
  \prod@fonttrue 
\fi

\def\newfam{\alloc@8\fam\chardef\sixt@@n} 

\ifprod@font
\font\fiverm=mtr10 at 5pt
\font\fivebf=mtbx10 at 5pt
\font\fiveit=mtti10 at 5pt
\font\fivesl=mtsl10 at 5pt
\font\fivett=mttt10 at 5pt     \hyphenchar\fivett=-1
\font\fivecsc=mtcsc10 at 5pt
\font\fivesf=mtss10 at 5pt
\font\fivei=mtmi10 at 5pt      \skewchar\fivei='177
\font\fivemib=mtmib10 at 5pt   \skewchar\fivemib='177
\font\fivesy=mtsy10 at 5pt     \skewchar\fivesy='60
\font\fivesyb=mtbsy10 at 5pt   \skewchar\fivesyb='60

\font\sixrm=mtr10 at 6pt
\font\sixbf=mtbx10 at 6pt
\font\sixit=mtti10 at 6pt
\font\sixsl=mtsl10 at 6pt
\font\sixtt=mttt10 at 6pt      \hyphenchar\sixtt=-1
\font\sixcsc=mtcsc10 at 6pt
\font\sixsf=mtss10 at 6pt
\font\sixi=mtmi10 at 6pt       \skewchar\sixi='177
\font\sixmib=mtmib10 at 6pt    \skewchar\sixmib='177
\font\sixsy=mtsy10 at 6pt      \skewchar\sixsy='60
\font\sixsyb=mtbsy10 at 6pt    \skewchar\sixsyb='60

\font\sevenrm=mtr10 at 7pt
\font\sevenbf=mtbx10 at 7pt
\font\sevenit=mtti10 at 7pt
\font\sevensl=mtsl10 at 7pt
\font\seventt=mttt10 at 7pt     \hyphenchar\seventt=-1
\font\sevencsc=mtcsc10 at 7pt
\font\sevensf=mtss10 at 7pt
\font\seveni=mtmi10 at 7pt      \skewchar\seveni='177
\font\sevenmib=mtmib10 at 7pt   \skewchar\sevenmib='177
\font\sevensy=mtsy10 at 7pt     \skewchar\sevensy='60
\font\sevensyb=mtbsy10 at 7pt   \skewchar\sevensyb='60

\font\eightrm=mtr10 at 8pt
\font\eightbf=mtbx10 at 8pt
\font\eightit=mtti10 at 8pt
\font\eighti=mtmi10 at 8pt      \skewchar\eighti='177
\font\eightmib=mtmib10 at 8pt   \skewchar\eightmib='177
\font\eightsy=mtsy10 at 8pt     \skewchar\eightsy='60
\font\eightsyb=mtbsy10 at 8pt   \skewchar\eightsyb='60
\font\eightsl=mtsl10 at 8pt
\font\eighttt=mttt10 at 8pt     \hyphenchar\eighttt=-1
\font\eightcsc=mtcsc10 at 8pt
\font\eightsf=mtss10 at 8pt

\font\ninerm=mtr10 at 9pt
\font\ninebf=mtbx10 at 9pt
\font\nineit=mtti10 at 9pt
\font\ninei=mtmi10 at 9pt      \skewchar\ninei='177
\font\ninemib=mtmib10 at 9pt   \skewchar\ninemib='177
\font\ninesy=mtsy10 at 9pt     \skewchar\ninesy='60
\font\ninesyb=mtbsy10 at 9pt   \skewchar\ninesyb='60
\font\ninesl=mtsl10 at 9pt
\font\ninett=mttt10 at 9pt     \hyphenchar\ninett=-1
\font\ninecsc=mtcsc10 at 9pt
\font\ninesf=mtss10 at 9pt

\font\tenrm=mtr10
\font\tenbf=mtbx10
\font\tenit=mtti10
\font\teni=mtmi10		\skewchar\teni='177
\font\tenmib=mtmib10	\skewchar\tenmib='177
\font\tensy=mtsy10		\skewchar\tensy='60
\font\tensyb=mtbsy10	\skewchar\tensyb='60
\font\tenex=cmex10
\font\tensl=mtsl10
\font\tentt=mttt10		\hyphenchar\tentt=-1
\font\tencsc=mtcsc10
\font\tensf=mtss10

\font\elevenrm=mtr10 at 11pt
\font\elevenbf=mtbx10 at 11pt
\font\elevenit=mtti10 at 11pt
\font\eleveni=mtmi10 at 11pt      \skewchar\eleveni='177
\font\elevenmib=mtmib10 at 11pt   \skewchar\elevenmib='177
\font\elevensy=mtsy10 at 11pt     \skewchar\elevensy='60
\font\elevensyb=mtbsy10 at 11pt   \skewchar\elevensyb='60
\font\elevensl=mtsl10 at 11pt
\font\eleventt=mttt10 at 11pt     \hyphenchar\eleventt=-1
\font\elevencsc=mtcsc10 at 11pt
\font\elevensf=mtss10 at 11pt

\font\twelverm=mtr10 at 12pt
\font\twelvebf=mtbx10 at 12pt
\font\twelveit=mtti10 at 12pt
\font\twelvesl=mtsl10 at 12pt
\font\twelvett=mttt10 at 12pt     \hyphenchar\twelvett=-1
\font\twelvecsc=mtcsc10 at 12pt
\font\twelvesf=mtss10 at 12pt
\font\twelvei=mtmi10 at 12pt      \skewchar\twelvei='177
\font\twelvemib=mtmib10 at 12pt   \skewchar\twelvemib='177
\font\twelvesy=mtsy10 at 12pt     \skewchar\twelvesy='60
\font\twelvesyb=mtbsy10 at 12pt   \skewchar\twelvesyb='60

\font\fourteenrm=mtr10 at 14pt
\font\fourteenbf=mtbx10 at 14pt
\font\fourteenit=mtti10 at 14pt
\font\fourteeni=mtmi10 at 14pt      \skewchar\fourteeni='177
\font\fourteenmib=mtmib10 at 14pt   \skewchar\fourteenmib='177
\font\fourteensy=mtsy10 at 14pt     \skewchar\fourteensy='60
\font\fourteensyb=mtbsy10 at 14pt   \skewchar\fourteensyb='60
\font\fourteensl=mtsl10 at 14pt
\font\fourteentt=mttt10 at 14pt     \hyphenchar\fourteentt=-1
\font\fourteencsc=mtcsc10 at 14pt
\font\fourteensf=mtss10 at 14pt

\font\seventeenrm=mtr10 at 17pt
\font\seventeenbf=mtbx10 at 17pt
\font\seventeenit=mtti10 at 17pt
\font\seventeeni=mtmi10 at 17pt      \skewchar\seventeeni='177
\font\seventeenmib=mtmib10 at 17pt   \skewchar\seventeenmib='177
\font\seventeensy=mtsy10 at 17pt     \skewchar\seventeensy='60
\font\seventeensyb=mtbsy10 at 17pt   \skewchar\seventeensyb='60
\font\seventeensl=mtsl10 at 17pt
\font\seventeentt=mttt10 at 17pt     \hyphenchar\seventeentt=-1
\font\seventeencsc=mtcsc10 at 17pt
\font\seventeensf=mtss10 at 17pt


\newfam\xmfam
\newfam\ymfam

\font\fivexm=mtxm10 at 5pt
\font\sixxm=mtxm10 at 6pt
\font\sevenxm=mtxm10 at 7pt
\font\eightxm=mtxm10 at 8pt
\font\ninexm=mtxm10 at 9pt
\font\tenxm=mtxm10
\font\elevenxm=mtxm10 at 11pt
\font\twelvexm=mtxm10 at 12pt
\font\fourteenxm=mtxm10 at 14pt
\font\seventeenxm=mtxm10 at 17pt

\font\fiveym=mtym10 at 5pt
\font\sixym=mtym10 at 6pt
\font\sevenym=mtym10 at 7pt
\font\eightym=mtym10 at 8pt
\font\nineym=mtym10 at 9pt
\font\tenym=mtym10
\font\elevenym=mtym10 at 11pt
\font\twelveym=mtym10 at 12pt
\font\fourteenym=mtym10 at 14pt
\font\seventeenym=mtym10 at 17pt
\else
\font\fiverm=cmr5
\font\fivei=cmmi5             \skewchar\fivei='177
\font\fivemib=cmmib10 at 5pt  \skewchar\fivemib='177
\font\fivesy=cmsy5            \skewchar\fivesy='60
\font\fivesyb=cmbsy10 at 5pt  \skewchar\fivesyb='60
\font\fivebf=cmbx5

\font\sixrm=cmr6
\font\sixi=cmmi6             \skewchar\sixi='177
\font\sixmib=cmmib10 at 6pt  \skewchar\sixmib='177
\font\sixsy=cmsy6            \skewchar\sixsy='60
\font\sixsyb=cmbsy10 at 6pt  \skewchar\sixsyb='60
\font\sixbf=cmbx6

\font\sevenrm=cmr7
\font\seveni=cmmi7             \skewchar\seveni='177
\font\sevenmib=cmmib10 at 7pt  \skewchar\sevenmib='177
\font\sevensy=cmsy7            \skewchar\sevensy='60
\font\sevensyb=cmbsy10 at 7pt  \skewchar\sevensyb='60
\font\sevenbf=cmbx7

\font\eightrm=cmr8
\font\eightbf=cmbx8
\font\eightit=cmti8
\font\eighti=cmmi8			\skewchar\eighti='177
\font\eightmib=cmmib10 at 8pt	\skewchar\eightmib='177
\font\eightsy=cmsy8			\skewchar\eightsy='60
\font\eightsyb=cmbsy10 at 8pt	\skewchar\eightsyb='60
\font\eightsl=cmsl8
\font\eighttt=cmtt8			\hyphenchar\eighttt=-1
\font\eightcsc=cmcsc10 at 8pt
\font\eightsf=cmss8

\font\ninerm=cmr9
\font\ninebf=cmbx9
\font\nineit=cmti9
\font\ninei=cmmi9			\skewchar\ninei='177
\font\ninemib=cmmib10 at 9pt	\skewchar\ninemib='177
\font\ninesy=cmsy9			\skewchar\ninesy='60
\font\ninesyb=cmbsy10 at 9pt	\skewchar\ninesyb='60
\font\ninesl=cmsl9
\font\ninett=cmtt9			\hyphenchar\ninett=-1
\font\ninecsc=cmcsc10 at 9pt
\font\ninesf=cmss9

\font\tenrm=cmr10
\font\tenbf=cmbx10
\font\tenit=cmti10
\font\teni=cmmi10		\skewchar\teni='177
\font\tenmib=cmmib10	\skewchar\tenmib='177
\font\tensy=cmsy10		\skewchar\tensy='60
\font\tensyb=cmbsy10	\skewchar\tensyb='60
\font\tenex=cmex10
\font\tensl=cmsl10
\font\tentt=cmtt10		\hyphenchar\tentt=-1
\font\tencsc=cmcsc10
\font\tensf=cmss10

\font\elevenrm=cmr10 scaled \magstephalf
\font\elevenbf=cmbx10 scaled \magstephalf
\font\elevenit=cmti10 scaled \magstephalf
\font\eleveni=cmmi10 scaled \magstephalf	\skewchar\eleveni='177
\font\elevenmib=cmmib10 scaled \magstephalf	\skewchar\elevenmib='177
\font\elevensy=cmsy10 scaled \magstephalf	\skewchar\elevensy='60
\font\elevensyb=cmbsy10 scaled \magstephalf	\skewchar\elevensyb='60
\font\elevensl=cmsl10 scaled \magstephalf
\font\eleventt=cmtt10 scaled \magstephalf	\hyphenchar\eleventt=-1
\font\elevencsc=cmcsc10 scaled \magstephalf
\font\elevensf=cmss10 scaled \magstephalf

\font\twelverm=cmr10 scaled \magstep1
\font\twelvebf=cmbx10 scaled \magstep1
\font\twelvei=cmmi10 scaled \magstep1      \skewchar\twelvei='177
\font\twelvemib=cmmib10 scaled \magstep1   \skewchar\twelvemib='177
\font\twelvesy=cmsy10 scaled \magstep1     \skewchar\twelvesy='60
\font\twelvesyb=cmbsy10 scaled \magstep1   \skewchar\twelvesyb='60

\font\fourteenrm=cmr10 scaled \magstep2
\font\fourteenbf=cmbx10 scaled \magstep2
\font\fourteenit=cmti10 scaled \magstep2
\font\fourteeni=cmmi10 scaled \magstep2		\skewchar\fourteeni='177
\font\fourteenmib=cmmib10 scaled \magstep2	\skewchar\fourteenmib='177
\font\fourteensy=cmsy10 scaled \magstep2	\skewchar\fourteensy='60
\font\fourteensyb=cmbsy10 scaled \magstep2	\skewchar\fourteensyb='60
\font\fourteensl=cmsl10 scaled \magstep2
\font\fourteentt=cmtt10 scaled \magstep2	\hyphenchar\fourteentt=-1
\font\fourteencsc=cmcsc10 scaled \magstep2
\font\fourteensf=cmss10 scaled \magstep2

\font\seventeenrm=cmr10 scaled \magstep3
\font\seventeenbf=cmbx10 scaled \magstep3
\font\seventeenit=cmti10 scaled \magstep3
\font\seventeeni=cmmi10 scaled \magstep3	\skewchar\seventeeni='177
\font\seventeenmib=cmmib10 scaled \magstep3	\skewchar\seventeenmib='177
\font\seventeensy=cmsy10 scaled \magstep3	\skewchar\seventeensy='60
\font\seventeensyb=cmbsy10 scaled \magstep3	\skewchar\seventeensyb='60
\font\seventeensl=cmsl10 scaled \magstep3
\font\seventeentt=cmtt10 scaled \magstep3	\hyphenchar\seventeentt=-1
\font\seventeencsc=cmcsc10 scaled \magstep3
\font\seventeensf=cmss10 scaled \magstep3
\fi

\def\hexnumber#1{\ifcase#1 0\or1\or2\or3\or4\or5\or6\or7\or8\or9\or
  A\or B\or C\or D\or E\or F\fi}

\ifprod@font
  \edef\@xm{\hexnumber\xmfam}
  \edef\@ym{\hexnumber\ymfam}
\fi

\def\makestrut{%
  \setbox\strutbox=\hbox{%
    \vrule height.7\baselineskip depth.3\baselineskip width \z@}%
}

\def\baselinestretch{1}
\newskip\tmp@bls

\def\b@ls#1{
  \tmp@bls=#1\relax
  \baselineskip=#1\relax\makestrut
  \normalbaselineskip=\baselinestretch\tmp@bls
  \normalbaselines
}

\def\nostb@ls#1{
  \normalbaselineskip=#1\relax
  \normalbaselines
  \makestrut
}

%

\newfam\mibfam 
\newfam\sybfam 
\newfam\scfam  
\newfam\sffam  

\def\mit{\fam\@ne}

\def\cal{\fam\tw@}

\def\em{\ifdim\fontdimen1\font>\z@ \rm\else\it\fi}

\textfont3=\tenex
\scriptfont3=\tenex
\scriptscriptfont3=\tenex

\setbox0=\hbox{\tenex B} \p@renwd=\wd0 

\def\eightpoint{
  \def\rm{\fam0\eightrm}%
  \textfont0=\eightrm \scriptfont0=\sixrm \scriptscriptfont0=\fiverm%
  \textfont1=\eighti  \scriptfont1=\sixi  \scriptscriptfont1=\fivei%
  \textfont2=\eightsy \scriptfont2=\sixsy \scriptscriptfont2=\fivesy%
  \textfont\itfam=\eightit\def\it{\fam\itfam\eightit}%
  \ifprod@font
    \scriptfont\itfam=\sixit
      \scriptscriptfont\itfam=\fiveit
  \else
    \scriptfont\itfam=\eightit
      \scriptscriptfont\itfam=\eightit
  \fi
  \textfont\bffam=\eightbf%
    \scriptfont\bffam=\sixbf%
      \scriptscriptfont\bffam=\fivebf%
  \def\bf{\fam\bffam\eightbf}%
  \textfont\slfam=\eightsl\def\sl{\fam\slfam\eightsl}%
  \ifprod@font
    \scriptfont\slfam=\sixsl
      \scriptscriptfont\slfam=\fivesl
  \else
    \scriptfont\slfam=\eightsl
      \scriptscriptfont\slfam=\eightsl
  \fi
  \textfont\ttfam=\eighttt\def\tt{\fam\ttfam\eighttt}%
  \ifprod@font
    \scriptfont\ttfam=\sixtt
      \scriptscriptfont\ttfam=\fivett
  \else
    \scriptfont\ttfam=\eighttt
      \scriptscriptfont\ttfam=\eighttt
  \fi
  \textfont\scfam=\eightcsc\def\sc{\fam\scfam\eightcsc}%
  \ifprod@font
    \scriptfont\scfam=\sixcsc
      \scriptscriptfont\scfam=\fivecsc
  \else
    \scriptfont\scfam=\eightcsc
      \scriptscriptfont\scfam=\eightcsc
  \fi
  \textfont\sffam=\eightsf\def\sf{\fam\sffam\eightsf}%
  \ifprod@font
    \scriptfont\sffam=\sixsf
      \scriptscriptfont\sffam=\fivesf
  \else
    \scriptfont\sffam=\eightsf
      \scriptscriptfont\sffam=\eightsf
  \fi
  \textfont\mibfam=\eightmib
    \scriptfont\mibfam=\sixmib
      \scriptscriptfont\mibfam=\fivemib
  \textfont\sybfam=\eightsyb
    \scriptfont\sybfam=\sixsyb
      \scriptscriptfont\sybfam=\fivesyb
  \ifprod@font
    \textfont\xmfam=\eightxm
      \scriptfont\xmfam=\sixxm
        \scriptscriptfont\xmfam=\fivexm
    \textfont\ymfam=\eightym
      \scriptfont\ymfam=\sixym
        \scriptscriptfont\ymfam=\fiveym
  \fi
  \def\oldstyle{\fam\@ne\eighti}%
  \def\boldstyle{\fam\mibfam\eightmib}%
  \b@ls{10pt}\rm%
}

\def\ninepoint{
  \def\rm{\fam0\ninerm}%
  \textfont0=\ninerm \scriptfont0=\sixrm \scriptscriptfont0=\fiverm%
  \textfont1=\ninei  \scriptfont1=\sixi  \scriptscriptfont1=\fivei%
  \textfont2=\ninesy \scriptfont2=\sixsy \scriptscriptfont2=\fivesy%
  \textfont\itfam=\nineit\def\it{\fam\itfam\nineit}%
  \ifprod@font
    \scriptfont\itfam=\sixit
      \scriptscriptfont\itfam=\fiveit
  \else
    \scriptfont\itfam=\nineit
      \scriptscriptfont\itfam=\nineit
  \fi
  \textfont\bffam=\ninebf%
    \scriptfont\bffam=\sixbf%
      \scriptscriptfont\bffam=\fivebf%
  \def\bf{\fam\bffam\ninebf}%
  \textfont\slfam=\ninesl\def\sl{\fam\slfam\ninesl}%
  \ifprod@font
    \scriptfont\slfam=\sixsl
      \scriptscriptfont\slfam=\fivesl
  \else
    \scriptfont\slfam=\ninesl
      \scriptscriptfont\slfam=\ninesl
  \fi
  \textfont\ttfam=\ninett\def\tt{\fam\ttfam\ninett}%
  \ifprod@font
    \scriptfont\ttfam=\sixtt
      \scriptscriptfont\ttfam=\fivett
  \else
    \scriptfont\ttfam=\ninett
      \scriptscriptfont\ttfam=\ninett
  \fi
  \textfont\scfam=\ninecsc\def\sc{\fam\scfam\ninecsc}%
  \ifprod@font
    \scriptfont\scfam=\sixcsc
      \scriptscriptfont\scfam=\fivecsc
  \else
    \scriptfont\scfam=\ninecsc
      \scriptscriptfont\scfam=\ninecsc
  \fi
  \textfont\sffam=\ninesf\def\sf{\fam\sffam\ninesf}%
  \ifprod@font
    \scriptfont\sffam=\sixsf
      \scriptscriptfont\sffam=\fivesf
  \else
    \scriptfont\sffam=\ninesf
      \scriptscriptfont\sffam=\ninesf
  \fi
  \textfont\mibfam=\ninemib
    \scriptfont\mibfam=\sixmib
      \scriptscriptfont\mibfam=\fivemib
  \textfont\sybfam=\ninesyb
    \scriptfont\sybfam=\sixsyb
      \scriptscriptfont\sybfam=\fivesyb
  \ifprod@font
    \textfont\xmfam=\ninexm
      \scriptfont\xmfam=\sixxm
        \scriptscriptfont\xmfam=\fivexm
    \textfont\ymfam=\nineym
      \scriptfont\ymfam=\sixym
        \scriptscriptfont\ymfam=\fiveym
  \fi
  \def\oldstyle{\fam\@ne\ninei}%
  \def\boldstyle{\fam\mibfam\ninemib}%
  \b@ls{\TextLeading plus \Feathering}\rm%
}

\def\tenpoint{
  \def\rm{\fam0\tenrm}%
  \textfont0=\tenrm \scriptfont0=\sevenrm \scriptscriptfont0=\fiverm%
  \textfont1=\teni  \scriptfont1=\seveni  \scriptscriptfont1=\fivei%
  \textfont2=\tensy \scriptfont2=\sevensy \scriptscriptfont2=\fivesy%
  \textfont\itfam=\tenit\def\it{\fam\itfam\tenit}%
  \ifprod@font
    \scriptfont\itfam=\sevenit
      \scriptscriptfont\itfam=\fiveit
  \else
    \scriptfont\itfam=\tenit
      \scriptscriptfont\itfam=\tenit
  \fi
  \textfont\bffam=\tenbf%
    \scriptfont\bffam=\sevenbf%
      \scriptscriptfont\bffam=\fivebf%
  \def\bf{\fam\bffam\tenbf}%
  \textfont\slfam=\tensl\def\sl{\fam\slfam\tensl}%
  \ifprod@font
    \scriptfont\slfam=\sevensl
      \scriptscriptfont\slfam=\fivesl
  \else
    \scriptfont\slfam=\tensl
      \scriptscriptfont\slfam=\tensl
  \fi
  \textfont\ttfam=\tentt\def\tt{\fam\ttfam\tentt}%
  \ifprod@font
    \scriptfont\ttfam=\seventt
      \scriptscriptfont\ttfam=\fivett
  \else
    \scriptfont\ttfam=\tentt
      \scriptscriptfont\ttfam=\tentt
  \fi
  \textfont\scfam=\tencsc\def\sc{\fam\scfam\tencsc}%
  \ifprod@font
    \scriptfont\scfam=\sevencsc
      \scriptscriptfont\scfam=\fivecsc
  \else
    \scriptfont\scfam=\tencsc
      \scriptscriptfont\scfam=\tencsc
  \fi
  \textfont\sffam=\tensf\def\sf{\fam\sffam\tensf}%
  \ifprod@font
    \scriptfont\sffam=\sevensf
      \scriptscriptfont\sffam=\fivesf
  \else
    \scriptfont\sffam=\tensf
      \scriptscriptfont\sffam=\tensf
  \fi
  \textfont\mibfam=\tenmib
    \scriptfont\mibfam=\sevenmib
      \scriptscriptfont\mibfam=\fivemib
  \textfont\sybfam=\tensyb
    \scriptfont\sybfam=\sevensyb
      \scriptscriptfont\sybfam=\fivesyb
  \ifprod@font
    \textfont\xmfam=\tenxm
      \scriptfont\xmfam=\sevenxm
        \scriptscriptfont\xmfam=\fivexm
    \textfont\ymfam=\tenym
      \scriptfont\ymfam=\sevenym
        \scriptscriptfont\ymfam=\fiveym
  \fi
  \def\oldstyle{\fam\@ne\teni}%
  \def\boldstyle{\fam\mibfam\tenmib}%
  \b@ls{11pt}\rm%
}

\def\elevenpoint{
  \def\rm{\fam0\elevenrm}%
  \textfont0=\elevenrm \scriptfont0=\eightrm \scriptscriptfont0=\sixrm%
  \textfont1=\eleveni  \scriptfont1=\eighti  \scriptscriptfont1=\sixi%
  \textfont2=\elevensy \scriptfont2=\eightsy \scriptscriptfont2=\sixsy%
  \textfont\itfam=\elevenit\def\it{\fam\itfam\elevenit}%
  \ifprod@font
    \scriptfont\itfam=\eightit
      \scriptscriptfont\itfam=\sixit
  \else
    \scriptfont\itfam=\elevenit
      \scriptscriptfont\itfam=\elevenit
  \fi
  \textfont\bffam=\elevenbf%
    \scriptfont\bffam=\eightbf%
      \scriptscriptfont\bffam=\sixbf%
  \def\bf{\fam\bffam\elevenbf}%
  \textfont\slfam=\elevensl\def\sl{\fam\slfam\elevensl}%
  \ifprod@font
    \scriptfont\slfam=\eightsl
      \scriptscriptfont\slfam=\sixsl
  \else
    \scriptfont\slfam=\elevensl
      \scriptscriptfont\slfam=\elevensl
  \fi
  \textfont\ttfam=\eleventt\def\tt{\fam\ttfam\eleventt}%
  \ifprod@font
    \scriptfont\ttfam=\eighttt
      \scriptscriptfont\ttfam=\sixtt
  \else
    \scriptfont\ttfam=\eleventt
      \scriptscriptfont\ttfam=\eleventt
  \fi
  \textfont\scfam=\elevencsc\def\sc{\fam\scfam\elevencsc}%
  \ifprod@font
    \scriptfont\scfam=\eightcsc
      \scriptscriptfont\scfam=\sixcsc
  \else
    \scriptfont\scfam=\elevencsc
      \scriptscriptfont\scfam=\elevencsc
  \fi
  \textfont\sffam=\elevensf\def\sf{\fam\sffam\elevensf}%
  \ifprod@font
    \scriptfont\sffam=\eightsf
      \scriptscriptfont\sffam=\sixsf
  \else
    \scriptfont\sffam=\elevensf
      \scriptscriptfont\sffam=\elevensf
  \fi
  \textfont\mibfam=\elevenmib
    \scriptfont\mibfam=\eightmib
      \scriptscriptfont\mibfam=\sixmib
  \textfont\sybfam=\elevensyb
    \scriptfont\sybfam=\eightsyb
      \scriptscriptfont\sybfam=\sixsyb
  \ifprod@font
    \textfont\xmfam=\elevenxm
      \scriptfont\xmfam=\eightxm
       \scriptscriptfont\xmfam=\sixxm
    \textfont\ymfam=\elevenym
      \scriptfont\ymfam=\eightym
        \scriptscriptfont\ymfam=\sixym
   \fi
  \def\oldstyle{\fam\@ne\eleveni}%
  \def\boldstyle{\fam\mibfam\elevenmib}%
  \b@ls{13pt}\rm%
}

\def\fourteenpoint{
  \def\rm{\fam0\fourteenrm}%
  \textfont0\fourteenrm  \scriptfont0\tenrm  \scriptscriptfont0\sevenrm%
  \textfont1\fourteeni   \scriptfont1\teni   \scriptscriptfont1\seveni%
  \textfont2\fourteensy  \scriptfont2\tensy  \scriptscriptfont2\sevensy%
  \textfont\itfam=\fourteenit\def\it{\fam\itfam\fourteenit}%
  \ifprod@font
    \scriptfont\itfam=\tenit
      \scriptscriptfont\itfam=\sevenit
  \else
    \scriptfont\itfam=\fourteenit
      \scriptscriptfont\itfam=\fourteenit
  \fi
  \textfont\bffam=\fourteenbf%
    \scriptfont\bffam=\tenbf%
      \scriptscriptfont\bffam=\sevenbf%
  \def\bf{\fam\bffam\fourteenbf}%
  \textfont\slfam=\fourteensl\def\sl{\fam\slfam\fourteensl}%
  \ifprod@font
    \scriptfont\slfam=\tensl
      \scriptscriptfont\slfam=\sevensl
  \else
    \scriptfont\slfam=\fourteensl
      \scriptscriptfont\slfam=\fourteensl
  \fi
  \textfont\ttfam=\fourteentt\def\tt{\fam\ttfam\fourteentt}%
  \ifprod@font
    \scriptfont\ttfam=\tentt
      \scriptscriptfont\ttfam=\seventt
  \else
    \scriptfont\ttfam=\fourteentt
      \scriptscriptfont\ttfam=\fourteentt
  \fi
  \textfont\scfam=\fourteencsc\def\sc{\fam\scfam\fourteencsc}%
  \ifprod@font
    \scriptfont\scfam=\tencsc
      \scriptscriptfont\scfam=\sevencsc
  \else
    \scriptfont\scfam=\fourteencsc
      \scriptscriptfont\scfam=\fourteencsc
  \fi
  \textfont\sffam=\fourteensf\def\sf{\fam\sffam\fourteensf}%
  \ifprod@font
    \scriptfont\sffam=\tensf
      \scriptscriptfont\sffam=\sevensf
  \else
    \scriptfont\sffam=\fourteensf
      \scriptscriptfont\sffam=\fourteensf
  \fi
  \textfont\mibfam=\fourteenmib
    \scriptfont\mibfam=\tenmib
      \scriptscriptfont\mibfam=\sevenmib
  \textfont\sybfam=\fourteensyb
    \scriptfont\sybfam=\tensyb
      \scriptscriptfont\sybfam=\sevensyb
  \ifprod@font
    \textfont\xmfam=\fourteenxm
      \scriptfont\xmfam=\tenxm
        \scriptscriptfont\xmfam=\sevenxm
   \textfont\ymfam=\fourteenym
      \scriptfont\ymfam=\tenym
        \scriptscriptfont\ymfam=\sevenym
  \fi
  \def\oldstyle{\fam\@ne\fourteeni}%
  \def\boldstyle{\fam\mibfam\fourteenmib}%
  \b@ls{17pt}\rm%
}

\def\seventeenpoint{
  \def\rm{\fam0\seventeenrm}%
  \textfont0\seventeenrm  \scriptfont0\twelverm  \scriptscriptfont0\tenrm%
  \textfont1\seventeeni   \scriptfont1\twelvei   \scriptscriptfont1\teni%
  \textfont2\seventeensy  \scriptfont2\twelvesy  \scriptscriptfont2\tensy%
  \textfont\itfam=\seventeenit\def\it{\fam\itfam\seventeenit}%
  \ifprod@font
    \scriptfont\itfam=\twelveit
      \scriptscriptfont\itfam=\tenit
  \else
    \scriptfont\itfam=\seventeenit
      \scriptscriptfont\itfam=\seventeenit
  \fi
  \textfont\bffam=\seventeenbf%
    \scriptfont\bffam=\twelvebf%
      \scriptscriptfont\bffam=\tenbf%
  \def\bf{\fam\bffam\seventeenbf}%
  \textfont\slfam=\seventeensl\def\sl{\fam\slfam\seventeensl}%
  \ifprod@font
    \scriptfont\slfam=\twelvesl
      \scriptscriptfont\slfam=\tensl
  \else
    \scriptfont\slfam=\seventeensl
      \scriptscriptfont\slfam=\seventeensl
  \fi
  \textfont\ttfam=\seventeentt\def\tt{\fam\ttfam\seventeentt}%
  \ifprod@font
    \scriptfont\ttfam=\twelvett
      \scriptscriptfont\ttfam=\tentt
  \else
    \scriptfont\ttfam=\seventeentt
      \scriptscriptfont\ttfam=\seventeentt
  \fi
  \textfont\scfam=\seventeencsc\def\sc{\fam\scfam\seventeencsc}%
  \ifprod@font
    \scriptfont\scfam=\twelvecsc
      \scriptscriptfont\scfam=\tencsc
  \else
    \scriptfont\scfam=\seventeencsc
      \scriptscriptfont\scfam=\seventeencsc
  \fi
  \textfont\sffam=\seventeensf\def\sf{\fam\sffam\seventeensf}%
  \ifprod@font
    \scriptfont\sffam=\twelvesf
      \scriptscriptfont\sffam=\tensf
  \else
    \scriptfont\sffam=\seventeensf
      \scriptscriptfont\sffam=\seventeensf
  \fi
  \textfont\mibfam=\seventeenmib
    \scriptfont\mibfam=\twelvemib
      \scriptscriptfont\mibfam=\tenmib
  \textfont\sybfam=\seventeensyb
    \scriptfont\sybfam=\twelvesyb
      \scriptscriptfont\sybfam=\tensyb
  \ifprod@font
    \textfont\xmfam=\seventeenxm
      \scriptfont\xmfam=\twelvexm
        \scriptscriptfont\xmfam=\tenxm
    \textfont\ymfam=\seventeenym
      \scriptfont\ymfam=\twelveym
        \scriptscriptfont\ymfam=\tenym
  \fi
  \def\oldstyle{\fam\@ne\seventeeni}%
  \def\boldstyle{\fam\mibfam\seventeenmib}%
  \b@ls{20pt}\rm%
}

\lineskip=1pt      \normallineskip=\lineskip
\lineskiplimit=\z@ \normallineskiplimit=\lineskiplimit



\def\la{\mathrel{\mathchoice {\vcenter{\offinterlineskip\halign{\hfil
$\displaystyle##$\hfil\cr<\cr\sim\cr}}}
{\vcenter{\offinterlineskip\halign{\hfil$\textstyle##$\hfil\cr
<\cr\sim\cr}}}
{\vcenter{\offinterlineskip\halign{\hfil$\scriptstyle##$\hfil\cr
<\cr\sim\cr}}}
{\vcenter{\offinterlineskip\halign{\hfil$\scriptscriptstyle##$\hfil\cr
<\cr\sim\cr}}}}}

\def\ga{\mathrel{\mathchoice {\vcenter{\offinterlineskip\halign{\hfil
$\displaystyle##$\hfil\cr>\cr\sim\cr}}}
{\vcenter{\offinterlineskip\halign{\hfil$\textstyle##$\hfil\cr
>\cr\sim\cr}}}
{\vcenter{\offinterlineskip\halign{\hfil$\scriptstyle##$\hfil\cr
>\cr\sim\cr}}}
{\vcenter{\offinterlineskip\halign{\hfil$\scriptscriptstyle##$\hfil\cr
>\cr\sim\cr}}}}}

\def\getsto{\mathrel{\mathchoice {\vcenter{\offinterlineskip
\halign{\hfil
$\displaystyle##$\hfil\cr\gets\cr\to\cr}}}
{\vcenter{\offinterlineskip\halign{\hfil$\textstyle##$\hfil\cr\gets
\cr\to\cr}}}
{\vcenter{\offinterlineskip\halign{\hfil$\scriptstyle##$\hfil\cr\gets
\cr\to\cr}}}
{\vcenter{\offinterlineskip\halign{\hfil$\scriptscriptstyle##$\hfil\cr
\gets\cr\to\cr}}}}}

\def\lid{\mathrel{\mathchoice {\vcenter{\offinterlineskip\halign{\hfil
$\displaystyle##$\hfil\cr<\cr\noalign{\vskip1.2pt}=\cr}}}
{\vcenter{\offinterlineskip\halign{\hfil$\textstyle##$\hfil\cr<\cr
\noalign{\vskip1.2pt}=\cr}}}
{\vcenter{\offinterlineskip\halign{\hfil$\scriptstyle##$\hfil\cr<\cr
\noalign{\vskip1pt}=\cr}}}
{\vcenter{\offinterlineskip\halign{\hfil$\scriptscriptstyle##$\hfil\cr
<\cr
\noalign{\vskip0.9pt}=\cr}}}}}

\def\gid{\mathrel{\mathchoice {\vcenter{\offinterlineskip\halign{\hfil
$\displaystyle##$\hfil\cr>\cr\noalign{\vskip1.2pt}=\cr}}}
{\vcenter{\offinterlineskip\halign{\hfil$\textstyle##$\hfil\cr>\cr
\noalign{\vskip1.2pt}=\cr}}}
{\vcenter{\offinterlineskip\halign{\hfil$\scriptstyle##$\hfil\cr>\cr
\noalign{\vskip1pt}=\cr}}}
{\vcenter{\offinterlineskip\halign{\hfil$\scriptscriptstyle##$\hfil\cr
>\cr
\noalign{\vskip0.9pt}=\cr}}}}}

\def\grole{\mathrel{\mathchoice {\vcenter{\offinterlineskip\halign{\hfil
$\displaystyle##$\hfil\cr>\cr\noalign{\vskip-1.5pt}<\cr}}}
{\vcenter{\offinterlineskip\halign{\hfil$\textstyle##$\hfil\cr
>\cr\noalign{\vskip-1.5pt}<\cr}}}
{\vcenter{\offinterlineskip\halign{\hfil$\scriptstyle##$\hfil\cr
>\cr\noalign{\vskip-1pt}<\cr}}}
{\vcenter{\offinterlineskip\halign{\hfil$\scriptscriptstyle##$\hfil\cr
>\cr\noalign{\vskip-0.5pt}<\cr}}}}}

\def\leogr{\mathrel{\mathchoice {\vcenter{\offinterlineskip\halign{\hfil
$\displaystyle##$\hfil\cr<\cr\noalign{\vskip-1.5pt}>\cr}}}
{\vcenter{\offinterlineskip\halign{\hfil$\textstyle##$\hfil\cr
<\cr\noalign{\vskip-1.5pt}>\cr}}}
{\vcenter{\offinterlineskip\halign{\hfil$\scriptstyle##$\hfil\cr
<\cr\noalign{\vskip-1pt}>\cr}}}
{\vcenter{\offinterlineskip\halign{\hfil$\scriptscriptstyle##$\hfil\cr
<\cr\noalign{\vskip-0.5pt}>\cr}}}}}

\def\loa{\mathrel{\mathchoice {\vcenter{\offinterlineskip\halign{\hfil
$\displaystyle##$\hfil\cr<\cr\approx\cr}}}
{\vcenter{\offinterlineskip\halign{\hfil$\textstyle##$\hfil\cr
<\cr\approx\cr}}}
{\vcenter{\offinterlineskip\halign{\hfil$\scriptstyle##$\hfil\cr
<\cr\approx\cr}}}
{\vcenter{\offinterlineskip\halign{\hfil$\scriptscriptstyle##$\hfil\cr
<\cr\approx\cr}}}}}

\def\goa{\mathrel{\mathchoice {\vcenter{\offinterlineskip\halign{\hfil
$\displaystyle##$\hfil\cr>\cr\approx\cr}}}
{\vcenter{\offinterlineskip\halign{\hfil$\textstyle##$\hfil\cr
>\cr\approx\cr}}}
{\vcenter{\offinterlineskip\halign{\hfil$\scriptstyle##$\hfil\cr
>\cr\approx\cr}}}
{\vcenter{\offinterlineskip\halign{\hfil$\scriptscriptstyle##$\hfil\cr
>\cr\approx\cr}}}}}

\def\diameter{{\ifmmode\mathchoice
{\ooalign{\hfil\hbox{$\displaystyle/$}\hfil\crcr
{\hbox{$\displaystyle\mathchar"20D$}}}}
{\ooalign{\hfil\hbox{$\textstyle/$}\hfil\crcr
{\hbox{$\textstyle\mathchar"20D$}}}}
{\ooalign{\hfil\hbox{$\scriptstyle/$}\hfil\crcr
{\hbox{$\scriptstyle\mathchar"20D$}}}}
{\ooalign{\hfil\hbox{$\scriptscriptstyle/$}\hfil\crcr
{\hbox{$\scriptscriptstyle\mathchar"20D$}}}}
\else{\ooalign{\hfil/\hfil\crcr\mathhexbox20D}}%
\fi}}

\def\sq{\ifmmode\squareforqed\else{\unskip\nobreak\hfil
\penalty50\hskip1em\null\nobreak\hfil\squareforqed
\parfillskip=0pt\finalhyphendemerits=0\endgraf}\fi}
\def\squareforqed{\hbox{\rlap{$\sqcap$}$\sqcup$}}


\def\bbbc{{\mathchoice {\setbox0=\hbox{$\displaystyle\rm C$}\hbox{\hbox
to0pt{\kern0.4\wd0\vrule height0.9\ht0\hss}\box0}}
{\setbox0=\hbox{$\textstyle\rm C$}\hbox{\hbox
to0pt{\kern0.4\wd0\vrule height0.9\ht0\hss}\box0}}
{\setbox0=\hbox{$\scriptstyle\rm C$}\hbox{\hbox
to0pt{\kern0.4\wd0\vrule height0.9\ht0\hss}\box0}}
{\setbox0=\hbox{$\scriptscriptstyle\rm C$}\hbox{\hbox
to0pt{\kern0.4\wd0\vrule height0.9\ht0\hss}\box0}}}}
\def\bbbq{{\mathchoice {\setbox0=\hbox{$\displaystyle\rm
Q$}\hbox{\raise
0.15\ht0\hbox to0pt{\kern0.4\wd0\vrule height0.8\ht0\hss}\box0}}
{\setbox0=\hbox{$\textstyle\rm Q$}\hbox{\raise
0.15\ht0\hbox to0pt{\kern0.4\wd0\vrule height0.8\ht0\hss}\box0}}
{\setbox0=\hbox{$\scriptstyle\rm Q$}\hbox{\raise
0.15\ht0\hbox to0pt{\kern0.4\wd0\vrule height0.7\ht0\hss}\box0}}
{\setbox0=\hbox{$\scriptscriptstyle\rm Q$}\hbox{\raise
0.15\ht0\hbox to0pt{\kern0.4\wd0\vrule height0.7\ht0\hss}\box0}}}}
\def\bbbt{{\mathchoice {\setbox0=\hbox{$\displaystyle\rm
T$}\hbox{\hbox to0pt{\kern0.3\wd0\vrule height0.9\ht0\hss}\box0}}
{\setbox0=\hbox{$\textstyle\rm T$}\hbox{\hbox
to0pt{\kern0.3\wd0\vrule height0.9\ht0\hss}\box0}}
{\setbox0=\hbox{$\scriptstyle\rm T$}\hbox{\hbox
to0pt{\kern0.3\wd0\vrule height0.9\ht0\hss}\box0}}
{\setbox0=\hbox{$\scriptscriptstyle\rm T$}\hbox{\hbox
to0pt{\kern0.3\wd0\vrule height0.9\ht0\hss}\box0}}}}
\def\bbbs{{\mathchoice
{\setbox0=\hbox{$\displaystyle     \rm S$}\hbox{\raise0.5\ht0\hbox
to0pt{\kern0.35\wd0\vrule height0.45\ht0\hss}\hbox
to0pt{\kern0.55\wd0\vrule height0.5\ht0\hss}\box0}}
{\setbox0=\hbox{$\textstyle        \rm S$}\hbox{\raise0.5\ht0\hbox
to0pt{\kern0.35\wd0\vrule height0.45\ht0\hss}\hbox
to0pt{\kern0.55\wd0\vrule height0.5\ht0\hss}\box0}}
{\setbox0=\hbox{$\scriptstyle      \rm S$}\hbox{\raise0.5\ht0\hbox
to0pt{\kern0.35\wd0\vrule height0.45\ht0\hss}\raise0.05\ht0\hbox
to0pt{\kern0.5\wd0\vrule height0.45\ht0\hss}\box0}}
{\setbox0=\hbox{$\scriptscriptstyle\rm S$}\hbox{\raise0.5\ht0\hbox
to0pt{\kern0.4\wd0\vrule height0.45\ht0\hss}\raise0.05\ht0\hbox
to0pt{\kern0.55\wd0\vrule height0.45\ht0\hss}\box0}}}}
\def\bbbz{{\mathchoice {\hbox{$\sf\textstyle Z\kern-0.4em Z$}}
{\hbox{$\sf\textstyle Z\kern-0.4em Z$}}
{\hbox{$\sf\scriptstyle Z\kern-0.3em Z$}}
{\hbox{$\sf\scriptscriptstyle Z\kern-0.2em Z$}}}}


\ifprod@font
  \mathchardef\la="3\@xm2E
  \mathchardef\getsto="3\@xm1C
  \mathchardef\lid="3\@xm35
  \mathchardef\grole="3\@xm3F
  \mathchardef\loa="3\@xm2F
  \mathchardef\ga="3\@xm26
  \mathchardef\gid="3\@xm3D
  \mathchardef\leogr="3\@xm37
  \mathchardef\goa="3\@xm27
  \mathchardef\sq="0\@xm03
%
%
\def\diameter{{%
  \ifmmode
    \mathchoice
    {\ooalign{\hfil\hbox{$\displaystyle/$}\hfil\crcr
    {\lower.2ex\hbox{$\displaystyle\mathchar"20D$}}}}%
    {\ooalign{\hfil\hbox{$\textstyle/$}\hfil\crcr
    {\lower.2ex\hbox{$\textstyle\mathchar"20D$}}}}%
    {\ooalign{\hfil\hbox{$\scriptstyle/$}\hfil\crcr
    {\lower.1ex\hbox{$\scriptstyle\mathchar"20D$}}}}%
    {\ooalign{\hfil\hbox{$\scriptscriptstyle/$}\hfil\crcr
    {\lower.1ex\hbox{$\scriptscriptstyle\mathchar"20D$}}}}%
  \else
    {\ooalign{\hfil/\hfil\crcr\lower.2ex\hbox{\mathhexbox20D}}}%
  \fi
}}
%
%

\def\bbbc{{\Bbb{C}}}
\def\bbbq{{\Bbb{Q}}}
\def\bbbt{{\Bbb{T}}}
\def\bbbs{{\Bbb{S}}}
\def\bbbz{{\Bbb{Z}}}
\fi


\ifprod@font
\mathchardef\boxdot="2\@xm00
\mathchardef\boxplus="2\@xm01
\mathchardef\boxtimes="2\@xm02
\mathchardef\square="0\@xm03
\mathchardef\blacksquare="0\@xm04
\mathchardef\centerdot="2\@xm05
\mathchardef\lozenge="0\@xm06
\mathchardef\blacklozenge="0\@xm07
\mathchardef\circlearrowright="3\@xm08
\mathchardef\circlearrowleft="3\@xm09
\mathchardef\rightleftharpoons="3\@xm0A
\mathchardef\leftrightharpoons="3\@xm0B
\mathchardef\boxminus="2\@xm0C
\mathchardef\Vdash="3\@xm0D
\mathchardef\Vvdash="3\@xm0E
\mathchardef\vDash="3\@xm0F
\mathchardef\twoheadrightarrow="3\@xm10
\mathchardef\twoheadleftarrow="3\@xm11
\mathchardef\leftleftarrows="3\@xm12
\mathchardef\rightrightarrows="3\@xm13
\mathchardef\upuparrows="3\@xm14
\mathchardef\downdownarrows="3\@xm15
\mathchardef\upharpoonright="3\@xm16

\mathchardef\downharpoonright="3\@xm17
\mathchardef\upharpoonleft="3\@xm18
\mathchardef\downharpoonleft="3\@xm19
\mathchardef\rightarrowtail="3\@xm1A
\mathchardef\leftarrowtail="3\@xm1B
\mathchardef\leftrightarrows="3\@xm1C
\mathchardef\rightleftarrows="3\@xm1D
\mathchardef\Lsh="3\@xm1E
\mathchardef\Rsh="3\@xm1F
\mathchardef\rightsquigarrow="3\@xm20
\mathchardef\leftrightsquigarrow="3\@xm21
\mathchardef\looparrowleft="3\@xm22
\mathchardef\looparrowright="3\@xm23
\mathchardef\circeq="3\@xm24
\mathchardef\succsim="3\@xm25
\mathchardef\gtrsim="3\@xm26
\mathchardef\gtrapprox="3\@xm27
\mathchardef\multimap="3\@xm28
\mathchardef\therefore="3\@xm29
\mathchardef\because="3\@xm2A
\mathchardef\doteqdot="3\@xm2B

\mathchardef\triangleq="3\@xm2C
\mathchardef\precsim="3\@xm2D
\mathchardef\lesssim="3\@xm2E
\mathchardef\lessapprox="3\@xm2F
\mathchardef\eqslantless="3\@xm30
\mathchardef\eqslantgtr="3\@xm31
\mathchardef\curlyeqprec="3\@xm32
\mathchardef\curlyeqsucc="3\@xm33
\mathchardef\preccurlyeq="3\@xm34
\mathchardef\leqq="3\@xm35
\mathchardef\leqslant="3\@xm36
\mathchardef\lessgtr="3\@xm37
\mathchardef\backprime="0\@xm38
\mathchardef\risingdotseq="3\@xm3A
\mathchardef\fallingdotseq="3\@xm3B
\mathchardef\succcurlyeq="3\@xm3C
\mathchardef\geqq="3\@xm3D
\mathchardef\geqslant="3\@xm3E
\mathchardef\gtrless="3\@xm3F
\mathchardef\sqsubset="3\@xm40
\mathchardef\sqsupset="3\@xm41
\mathchardef\vartriangleright="3\@xm42
\mathchardef\vartriangleleft="3\@xm43
\mathchardef\trianglerighteq="3\@xm44
\mathchardef\trianglelefteq="3\@xm45
\mathchardef\bigstar="0\@xm46
\mathchardef\between="3\@xm47
\mathchardef\blacktriangledown="0\@xm48
\mathchardef\blacktriangleright="3\@xm49
\mathchardef\blacktriangleleft="3\@xm4A
\mathchardef\vartriangle="0\@xm4D
\mathchardef\blacktriangle="0\@xm4E
\mathchardef\triangledown="0\@xm4F
\mathchardef\eqcirc="3\@xm50
\mathchardef\lesseqgtr="3\@xm51
\mathchardef\gtreqless="3\@xm52
\mathchardef\lesseqqgtr="3\@xm53
\mathchardef\gtreqqless="3\@xm54
\mathchardef\Rrightarrow="3\@xm56
\mathchardef\Lleftarrow="3\@xm57
\mathchardef\veebar="2\@xm59
\mathchardef\barwedge="2\@xm5A
\mathchardef\doublebarwedge="2\@xm5B
\mathchardef\angle="0\@xm5C
\mathchardef\measuredangle="0\@xm5D
\mathchardef\sphericalangle="0\@xm5E
\mathchardef\varpropto="3\@xm5F
\mathchardef\smallsmile="3\@xm60
\mathchardef\smallfrown="3\@xm61
\mathchardef\Subset="3\@xm62
\mathchardef\Supset="3\@xm63
\mathchardef\Cup="2\@xm64

\mathchardef\Cap="2\@xm65

\mathchardef\curlywedge="2\@xm66
\mathchardef\curlyvee="2\@xm67
\mathchardef\leftthreetimes="2\@xm68
\mathchardef\rightthreetimes="2\@xm69
\mathchardef\subseteqq="3\@xm6A
\mathchardef\supseteqq="3\@xm6B
\mathchardef\bumpeq="3\@xm6C
\mathchardef\Bumpeq="3\@xm6D
\mathchardef\lll="3\@xm6E

\mathchardef\ggg="3\@xm6F

\mathchardef\circledS="0\@xm73
\mathchardef\pitchfork="3\@xm74
\mathchardef\dotplus="2\@xm75
\mathchardef\backsim="3\@xm76
\mathchardef\backsimeq="3\@xm77
\mathchardef\complement="0\@xm7B
\mathchardef\intercal="2\@xm7C
\mathchardef\circledcirc="2\@xm7D
\mathchardef\circledast="2\@xm7E
\mathchardef\circleddash="2\@xm7F
\def\ulcorner{\delimiter"4\@xm70\@xm70 }
\def\urcorner{\delimiter"5\@xm71\@xm71 }
\def\llcorner{\delimiter"4\@xm78\@xm78 }
\def\lrcorner{\delimiter"5\@xm79\@xm79 }
\def\yen{\mathhexbox\@xm55 }
\def\checkmark{\mathhexbox\@xm58 }
\def\circledR{\mathhexbox\@xm72 }
\def\maltese{\mathhexbox\@xm7A }
\mathchardef\lvertneqq="3\@ym00
\mathchardef\gvertneqq="3\@ym01
\mathchardef\nleq="3\@ym02
\mathchardef\ngeq="3\@ym03
\mathchardef\nless="3\@ym04
\mathchardef\ngtr="3\@ym05
\mathchardef\nprec="3\@ym06
\mathchardef\nsucc="3\@ym07
\mathchardef\lneqq="3\@ym08
\mathchardef\gneqq="3\@ym09
\mathchardef\nleqslant="3\@ym0A
\mathchardef\ngeqslant="3\@ym0B
\mathchardef\lneq="3\@ym0C
\mathchardef\gneq="3\@ym0D
\mathchardef\npreceq="3\@ym0E
\mathchardef\nsucceq="3\@ym0F
\mathchardef\precnsim="3\@ym10
\mathchardef\succnsim="3\@ym11
\mathchardef\lnsim="3\@ym12
\mathchardef\gnsim="3\@ym13
\mathchardef\nleqq="3\@ym14
\mathchardef\ngeqq="3\@ym15
\mathchardef\precneqq="3\@ym16
\mathchardef\succneqq="3\@ym17
\mathchardef\precnapprox="3\@ym18
\mathchardef\succnapprox="3\@ym19
\mathchardef\lnapprox="3\@ym1A
\mathchardef\gnapprox="3\@ym1B
\mathchardef\nsim="3\@ym1C
\mathchardef\ncong="3\@ym1D

\mathchardef\varsubsetneq="3\@ym20
\mathchardef\varsupsetneq="3\@ym21
\mathchardef\nsubseteqq="3\@ym22
\mathchardef\nsupseteqq="3\@ym23
\mathchardef\subsetneqq="3\@ym24
\mathchardef\supsetneqq="3\@ym25
\mathchardef\varsubsetneqq="3\@ym26
\mathchardef\varsupsetneqq="3\@ym27
\mathchardef\subsetneq="3\@ym28
\mathchardef\supsetneq="3\@ym29
\mathchardef\nsubseteq="3\@ym2A
\mathchardef\nsupseteq="3\@ym2B
\mathchardef\nparallel="3\@ym2C
\mathchardef\nmid="3\@ym2D
\mathchardef\nshortmid="3\@ym2E
\mathchardef\nshortparallel="3\@ym2F
\mathchardef\nvdash="3\@ym30
\mathchardef\nVdash="3\@ym31
\mathchardef\nvDash="3\@ym32
\mathchardef\nVDash="3\@ym33
\mathchardef\ntrianglerighteq="3\@ym34
\mathchardef\ntrianglelefteq="3\@ym35
\mathchardef\ntriangleleft="3\@ym36
\mathchardef\ntriangleright="3\@ym37
\mathchardef\nleftarrow="3\@ym38
\mathchardef\nrightarrow="3\@ym39
\mathchardef\nLeftarrow="3\@ym3A
\mathchardef\nRightarrow="3\@ym3B
\mathchardef\nLeftrightarrow="3\@ym3C
\mathchardef\nleftrightarrow="3\@ym3D
\mathchardef\divideontimes="2\@ym3E
\mathchardef\varnothing="0\@ym3F
\mathchardef\nexists="0\@ym40
\mathchardef\mho="0\@ym66
\mathchardef\eth="0\@ym67
\mathchardef\eqsim="3\@ym68
\mathchardef\beth="0\@ym69
\mathchardef\gimel="0\@ym6A
\mathchardef\daleth="0\@ym6B
\mathchardef\lessdot="3\@ym6C
\mathchardef\gtrdot="3\@ym6D
\mathchardef\ltimes="2\@ym6E
\mathchardef\rtimes="2\@ym6F
\mathchardef\shortmid="3\@ym70
\mathchardef\shortparallel="3\@ym71
\mathchardef\smallsetminus="2\@ym72
\mathchardef\thicksim="3\@ym73
\mathchardef\thickapprox="3\@ym74
\mathchardef\approxeq="3\@ym75
\mathchardef\succapprox="3\@ym76
\mathchardef\precapprox="3\@ym77
\mathchardef\curvearrowleft="3\@ym78
\mathchardef\curvearrowright="3\@ym79
\mathchardef\digamma="0\@ym7A
\mathchardef\varkappa="0\@ym7B
\mathchardef\hslash="0\@ym7D
\mathchardef\hbar="0\@ym7E
\mathchardef\backepsilon="3\@ym7F


\def\Bbb{\ifmmode\let\next\Bbb@\else
\def\next{\errmessage{Use \string\Bbb\space only in math mode}}\fi\next}
\def\Bbb@#1{{\Bbb@@{#1}}}
\def\Bbb@@#1{\fam\ymfam#1}
\fi


\def\Nulle{0} 
\def\Afe{1}   
\def\Hae{2}   
\def\Hbe{3}   
\def\Hce{4}   
\def\Hde{5}   


\newcount\LastMac       \LastMac=\Nulle

\newskip\half      \half=5.5pt plus 1.5pt minus 2.25pt
\newskip\one       \one=11pt plus 3pt minus 5.5pt
\newskip\onehalf   \onehalf=16.5pt plus 5.5pt minus 8.25pt
\newskip\two       \two=22pt plus 5.5pt minus 11pt

\def\Half{\addvspace{\half}}
\def\One{\addvspace{\one}}
\def\OneHalf{\addvspace{\onehalf}}
\def\Two{\addvspace{\two}}


\def\Raggedright{
  \rightskip=\z@ plus \hsize\relax
}

\def\Fullout{
  \rightskip=\z@\relax
}

\def\Hang#1#2{
  \hangindent=#1%
  \hangafter=#2\relax
}


\newif\ifsp@page
\def\pagestyle#1{\csname ps@#1\endcsname}
\def\thispagestyle#1{\global\sp@pagetrue\gdef\sp@type{#1}}

\def\ps@titlepage{%
  \def\@oddhead{\eightpoint\noindent \the\CatchLine
    \ifprod@font\else\qquad Printed\ \today\fi \hfil}%
  \let\@evenhead=\@oddhead
}

\def\ps@headings{%
  \def\@oddhead{\elevenpoint\it\noindent
    \hfill\the\RightHeader\hskip1.5em\rm\folio}%
  \def\@evenhead{\elevenpoint\noindent
    \folio\hskip1.5em\it\the\LeftHeader\hfill}%
}

\def\ps@plate{%
  \def\@oddhead{\eightpoint\noindent\plt@cap\hfil}%
  \def\@evenhead{\eightpoint\noindent\plt@cap\hfil}%
}



\def\title#1{
  \bgroup
    \vbox to 8pt{\vss}%
    \seventeenpoint
    \Raggedright
    \noindent \strut{\bf #1}\par
  \egroup
}

\def\author#1{
  \bgroup
    \ifnum\LastMac=\Afe \OneHalf\else \vskip 21pt\fi
    \fourteenpoint
    \Raggedright
    \noindent \strut #1\par
    \vskip 3pt%
  \egroup
}

\def\affiliation#1{
  \bgroup
    \vskip -4pt%
    \eightpoint
    \Raggedright
    \noindent \strut {\it #1}\par
  \egroup
  \LastMac=\Afe\relax
}

\def\acceptedline#1{
  \bgroup
    \Two
    \eightpoint
    \Raggedright
    \noindent \strut #1\par
  \egroup
}

\long\def\abstract#1{%
  \bgroup
    \vskip 20pt%
    \everypar{\Hang{11pc}{0}}%
    \noindent{\ninebf ABSTRACT}\par
    \tenpoint
    \Fullout
    \noindent #1\par
  \egroup
}

\long\def\keywords#1{
  \bgroup
    \Half
    \everypar{\Hang{11pc}{0}}%
    \tenpoint
    \Fullout
    \noindent\hbox{\bf Key words:}\ #1\par
  \egroup
}


\def\maketitle{%
  \EndOpening
  \ifsinglecol \else \MakePage\fi
}



\newif\ifAutoNumber \AutoNumberfalse
\newcount\Sec        
\newcount\SecSec
\newcount\SecSecSec

\Sec=\z@

\def\:{\let\@sptoken= } \:  
\def\:{\@xifnch} \expandafter\def\: {\futurelet\@tempc\@ifnch}

\def\@ifnextchar#1#2#3{%
  \let\@tempMACe #1%
  \def\@tempMACa{#2}%
  \def\@tempMACb{#3}%
  \futurelet \@tempMACc\@ifnch%
}

\def\@ifnch{%
\ifx \@tempMACc \@sptoken%
  \let\@tempMACd\@xifnch%
\else%
  \ifx \@tempMACc \@tempMACe%
    \let\@tempMACd\@tempMACa%
  \else%
    \let\@tempMACd\@tempMACb%
  \fi%
\fi%
\@tempMACd%
}

\def\@ifstar#1#2{\@ifnextchar *{\def\@tempMACa*{#1}\@tempMACa}{#2}}

\newskip\@tempskipb

\def\addvspace#1{%
  \ifvmode\else \endgraf\fi%
  \ifdim\lastskip=\z@%
    \vskip #1\relax%
  \else%
    \@tempskipb#1\relax\@xaddvskip%
  \fi%
}

\def\@xaddvskip{%
  \ifdim\lastskip<\@tempskipb%
    \vskip-\lastskip%
    \vskip\@tempskipb\relax%
  \else%
    \ifdim\@tempskipb<\z@%
      \ifdim\lastskip<\z@ \else%
        \advance\@tempskipb\lastskip%
        \vskip-\lastskip\vskip\@tempskipb%
      \fi%
    \fi%
  \fi%
}

\newskip\@tmpSKIP

\def\addpen#1{%
  \ifvmode
    \if@nobreak
    \else
      \ifdim\lastskip=\z@
        \penalty#1\relax
      \else
        \@tmpSKIP=\lastskip
        \vskip -\lastskip
        \penalty#1\vskip\@tmpSKIP
      \fi
    \fi
  \fi
}

\newcount\@clubpen   \@clubpen=\clubpenalty
\newif\if@nobreak    \@nobreakfalse

\def\@noafterindent{%
  \global\@nobreaktrue
  \everypar{\if@nobreak
              \global\@nobreakfalse
              \clubpenalty \@M
              {\setbox\z@\lastbox}%
              \LastMac=\Nulle\relax%
            \else
              \clubpenalty \@clubpen
              \everypar{}%
            \fi}
}

\newcount\gds@cbrk   \gds@cbrk=-300

\def\@nohdbrk{\interlinepenalty \@M\relax}

\let\@par=\par
\def\@restorepar{\def\par{\@par}}

\newif\if@endpe   \@endpefalse
 
\def\@doendpe{\@endpetrue \@nobreakfalse \LastMac=\Nulle\relax%
     \def\par{\@restorepar\everypar{}\par\@endpefalse}%
              \everypar{\setbox\z@\lastbox\everypar{}\@endpefalse}%
}

\def\section{\@ifstar{\@ssection}{\@section}}

\def\@section#1{
  \if@nobreak
    \everypar{}%
    \ifnum\LastMac=\Hae \addvspace{\half}\fi
  \else
    \addpen{\gds@cbrk}%
    \addvspace{\two}%
  \fi
  \bgroup
    \ninepoint\bf
    \Raggedright
    \ifAutoNumber
      \global\advance\Sec \@ne
      \noindent\@nohdbrk\number\Sec\hskip 1pc \uppercase{#1}\par
      \global\SecSec=\z@
    \else
      \noindent\@nohdbrk\uppercase{#1}\par
    \fi
  \egroup
  \nobreak
  \vskip\half
  \nobreak
  \@noafterindent
  \LastMac=\Hae\relax
}

\def\@ssection#1{
  \if@nobreak
    \everypar{}%
    \ifnum\LastMac=\Hae \addvspace{\half}\fi
  \else
    \addpen{\gds@cbrk}%
    \addvspace{\two}%
  \fi
  \bgroup
    \ninepoint\bf
    \Raggedright
    \noindent\@nohdbrk\uppercase{#1}\par
  \egroup
  \nobreak
  \vskip\half
  \nobreak
  \@noafterindent
  \LastMac=\Hae\relax
}

\def\subsection#1{
  \if@nobreak
    \everypar{}%
    \ifnum\LastMac=\Hae \addvspace{1pt plus 1pt minus .5pt}\fi
  \else
    \addpen{\gds@cbrk}%
    \addvspace{\onehalf}%
  \fi
  \bgroup
    \ninepoint\bf
    \Raggedright
    \ifAutoNumber
      \global\advance\SecSec \@ne
      \noindent\@nohdbrk\number\Sec.\number\SecSec \hskip 1pc\relax #1\par
      \global\SecSecSec=\z@
    \else
      \noindent\@nohdbrk #1\par
    \fi
  \egroup
  \nobreak
  \vskip\half
  \nobreak
  \@noafterindent
  \LastMac=\Hbe\relax
}

\def\subsubsection#1{
  \if@nobreak
    \everypar{}%
    \ifnum\LastMac=\Hbe \addvspace{1pt plus 1pt minus .5pt}\fi
  \else
    \addpen{\gds@cbrk}%
    \addvspace{\onehalf}%
  \fi
  \bgroup
    \ninepoint\it
    \Raggedright
    \ifAutoNumber
      \global\advance\SecSecSec \@ne
      \noindent\@nohdbrk\number\Sec.\number\SecSec.\number\SecSecSec
        \hskip 1pc\relax #1\par
    \else
      \noindent\@nohdbrk #1\par
    \fi
  \egroup
  \nobreak
  \vskip\half
  \nobreak
  \@noafterindent
  \LastMac=\Hce\relax
}

\def\paragraph#1{
  \if@nobreak
    \everypar{}%
  \else
    \addpen{\gds@cbrk}%
    \addvspace{\one}%
  \fi%
  \bgroup%
    \ninepoint\it
    \noindent #1\ \nobreak%
  \egroup
  \LastMac=\Hde\relax
  \ignorespaces
}




\def\beginlist{%
  \par\if@nobreak \else\addvspace{\half}\fi%
  \bgroup%
    \ninepoint
    \let\item=\list@item%
}

\def\list@item{%
  \par\noindent\hskip 1em\relax%
  \ignorespaces%
}

\def\endlist{\par\egroup\addvspace{\half}\@doendpe}


\def\beginrefs{%
  \par
  \bgroup
    \eightpoint
    \Raggedright
    \let\bibitem=\bib@item
}

\def\bib@item{%
  \par\parindent=1.5em\Hang{1.5em}{1}%
  \everypar={\Hang{1.5em}{1}\ignorespaces}%
  \noindent\ignorespaces
}

\def\endrefs{\par\egroup\@doendpe}


\newtoks\CatchLine

\def\@journal{Mon.\ Not.\ R.\ Astron.\ Soc.\ }  
\def\@pubyear{1996}        
\def\@pagerange{000--000}  
\def\@volume{000}          
\def\@microfiche{}         %

\def\pubyear#1{\gdef\@pubyear{#1}\@makecatchline}
\def\pagerange#1{\gdef\@pagerange{#1}\@makecatchline}
\def\volume#1{\gdef\@volume{#1}\@makecatchline}
\def\microfiche#1{\gdef\@microfiche{and Microfiche\ #1}\@makecatchline}

\def\@makecatchline{%
  \global\CatchLine{%
    {\rm \@journal {\bf \@volume},\ \@pagerange\ (\@pubyear)\ \@microfiche}}%
}

\@makecatchline 

\newtoks\LeftHeader
\def\shortauthor#1{
  \global\LeftHeader{#1}%
}

\newtoks\RightHeader
\def\shorttitle#1{
  \global\RightHeader{#1}%
}

\def\PageHead{
  \begingroup
    \ifsp@page
      \csname ps@\sp@type\endcsname
      \global\sp@pagefalse
    \fi
    \ifodd\pageno
      \let\the@head=\@oddhead
    \else
      \let\the@head=\@evenhead
    \fi
    \vbox to \z@{\vskip-22.5\p@%
      \hbox to \PageWidth{\vbox to8.5\p@{}%
        \the@head
      }%
    \vss}%
  \endgroup
  \nointerlineskip
}

\def\today{%
  \number\day\space
  \ifcase\month\or January\or February\or March\or April\or May\or June\or
    July\or August\or September\or October\or November\or December\fi
  \space\number\year%
}

\def\PageFoot{} 

\def\authorcomment#1{%
  \gdef\PageFoot{%
    \nointerlineskip%
    \vbox to 22pt{\vfil%
      \hbox to \PageWidth{\elevenpoint\noindent \hfil #1 \hfil}}%
  }%
}


\newif\ifplate@page
\newbox\plt@box

\def\beginplatepage{%
  \let\plate=\plate@head
  \let\caption=\fig@caption
  \global\setbox\plt@box=\vbox\bgroup
  \TEMPDIMEN=\PageWidth 
  \hsize=\PageWidth\relax
}

\def\endplatepage{\par\egroup\global\plate@pagetrue}
\def\plate@head#1{\gdef\plt@cap{#1}}


\def\letters{%
  \gdef\folio{\ifnum\pageno<\z@ L\romannumeral-\pageno
    \else L\number\pageno \fi}%
}


\everydisplay{\displaysetup}

\newif\ifeqno
\newif\ifleqno

\def\displaysetup#1$${%
 \displaytest#1\eqno\eqno\displaytest
}

\def\displaytest#1\eqno#2\eqno#3\displaytest{%
 \if!#3!\ldisplaytest#1\leqno\leqno\ldisplaytest
 \else\eqnotrue\leqnofalse\def\eqn{#2}\def\eq{#1}\fi
 \generaldisplay$$}

\def\ldisplaytest#1\leqno#2\leqno#3\ldisplaytest{%
 \def\eq{#1}%
 \if!#3!\eqnofalse\else\eqnotrue\leqnotrue
  \def\eqn{#2}\fi}

\def\generaldisplay{%
\ifeqno \ifleqno 
   \hbox to \hsize{\noindent
     $\displaystyle\eq$\hfil$\displaystyle\eqn$}
  \else
    \hbox to \hsize{\noindent
     $\displaystyle\eq$\hfil$\displaystyle\eqn$}
  \fi
 \else
 \hbox to \hsize{\vbox{\noindent
  $\displaystyle\eq$\hfil}}
 \fi
}


\def\@notice{%
  \par\Two%
  \noindent{\b@ls{11pt}\ninerm This paper has been produced using the
    Blackwell Scientific Publications \TeX\ macros.\par}%
}

\outer\def\bye{\@notice\par\vfill\supereject\end}


\def\start@mess{%
  Monthly notices of the RAS journal style (\@typeface)\space
    v\@version,\space \@verdate.%
}

\everyjob{\Warn{\start@mess}}



\newif\if@debug \@debugfalse  

\def\Print#1{\if@debug\immediate\write16{#1}\else \fi}
\def\Warn#1{\immediate\write16{#1}}
\def\wlog#1{}

\newcount\Iteration 

\def\Single{0} \def\Double{1}                 
\def\Figure{0} \def\Table{1}                  

\def\InStack{0}  
\def\InZoneA{1}
\def\InZoneB{2}
\def\InZoneC{3}

\newcount\TEMPCOUNT 
\newdimen\TEMPDIMEN 
\newbox\TEMPBOX     
\newbox\VOIDBOX     

\newcount\LengthOfStack 
\newcount\MaxItems      
\newcount\StackPointer
\newcount\Point         
\newcount\NextFigure    
\newcount\NextTable     
\newcount\NextItem      

\newcount\StatusStack   
\newcount\NumStack      
\newcount\TypeStack     
\newcount\SpanStack     
\newcount\BoxStack      

\newcount\ItemSTATUS    
\newcount\ItemNUMBER    
\newcount\ItemTYPE      
\newcount\ItemSPAN      
\newbox\ItemBOX         
\newdimen\ItemSIZE      

\newdimen\PageHeight    
\newdimen\TextLeading   
\newdimen\Feathering    
\newcount\LinesPerPage  
\newdimen\ColumnWidth   
\newdimen\ColumnGap     
\newdimen\PageWidth     
\newdimen\BodgeHeight   
\newcount\Leading       

\newdimen\ZoneBSize  
\newdimen\TextSize   
\newbox\ZoneABOX     
\newbox\ZoneBBOX     
\newbox\ZoneCBOX     

\newif\ifFirstSingleItem
\newif\ifFirstZoneA
\newif\ifMakePageInComplete
\newif\ifMoreFigures \MoreFiguresfalse 
\newif\ifMoreTables  \MoreTablesfalse  

\newif\ifFigInZoneB 
\newif\ifFigInZoneC 
\newif\ifTabInZoneB 
\newif\ifTabInZoneC

\newif\ifZoneAFullPage

\newbox\MidBOX    
\newbox\LeftBOX
\newbox\RightBOX
\newbox\PageBOX   

\newif\ifLeftCOL  
\LeftCOLtrue

\newdimen\ZoneBAdjust

\newcount\ItemFits
\def\Yes{1}
\def\No{2}


\MaxItems=15
\NextFigure=\z@        
\NextTable=\@ne

\BodgeHeight=6pt
\TextLeading=11pt    
\Leading=11
\Feathering=\z@      
\LinesPerPage=61     
\topskip=\TextLeading
\ColumnWidth=20pc    
\ColumnGap=2pc       

\newskip\ItemSepamount  
\ItemSepamount=\TextLeading plus \TextLeading minus 4pt

\parskip=\z@ plus .1pt
\parindent=18pt
\widowpenalty=\z@
\clubpenalty=10000
\tolerance=1500
\hbadness=1500
\abovedisplayskip=6pt plus 2pt minus 2pt
\belowdisplayskip=6pt plus 2pt minus 2pt
\abovedisplayshortskip=6pt plus 2pt minus 2pt
\belowdisplayshortskip=6pt plus 2pt minus 2pt

\ninepoint 


\PageHeight=682pt

\PageWidth=2\ColumnWidth
\advance\PageWidth by \ColumnGap

\pagestyle{headings}




\newcount\DUMMY \StatusStack=\allocationnumber
\newcount\DUMMY \newcount\DUMMY \newcount\DUMMY 
\newcount\DUMMY \newcount\DUMMY \newcount\DUMMY 
\newcount\DUMMY \newcount\DUMMY \newcount\DUMMY
\newcount\DUMMY \newcount\DUMMY \newcount\DUMMY 
\newcount\DUMMY \newcount\DUMMY \newcount\DUMMY

\newcount\DUMMY \NumStack=\allocationnumber
\newcount\DUMMY \newcount\DUMMY \newcount\DUMMY 
\newcount\DUMMY \newcount\DUMMY \newcount\DUMMY 
\newcount\DUMMY \newcount\DUMMY \newcount\DUMMY 
\newcount\DUMMY \newcount\DUMMY \newcount\DUMMY 
\newcount\DUMMY \newcount\DUMMY \newcount\DUMMY

\newcount\DUMMY \TypeStack=\allocationnumber
\newcount\DUMMY \newcount\DUMMY \newcount\DUMMY 
\newcount\DUMMY \newcount\DUMMY \newcount\DUMMY 
\newcount\DUMMY \newcount\DUMMY \newcount\DUMMY 
\newcount\DUMMY \newcount\DUMMY \newcount\DUMMY 
\newcount\DUMMY \newcount\DUMMY \newcount\DUMMY

\newcount\DUMMY \SpanStack=\allocationnumber
\newcount\DUMMY \newcount\DUMMY \newcount\DUMMY 
\newcount\DUMMY \newcount\DUMMY \newcount\DUMMY 
\newcount\DUMMY \newcount\DUMMY \newcount\DUMMY 
\newcount\DUMMY \newcount\DUMMY \newcount\DUMMY 
\newcount\DUMMY \newcount\DUMMY \newcount\DUMMY

\newbox\DUMMY   \BoxStack=\allocationnumber
\newbox\DUMMY   \newbox\DUMMY \newbox\DUMMY 
\newbox\DUMMY   \newbox\DUMMY \newbox\DUMMY 
\newbox\DUMMY   \newbox\DUMMY \newbox\DUMMY 
\newbox\DUMMY   \newbox\DUMMY \newbox\DUMMY 
\newbox\DUMMY   \newbox\DUMMY \newbox\DUMMY

\def\wlog{\immediate\write\m@ne}


\def\GetItemAll#1{%
 \GetItemSTATUS{#1}
 \GetItemNUMBER{#1}
 \GetItemTYPE{#1}
 \GetItemSPAN{#1}
 \GetItemBOX{#1}
}

\def\GetItemSTATUS#1{%
 \Point=\StatusStack
 \advance\Point by #1
 \global\ItemSTATUS=\count\Point
}

\def\GetItemNUMBER#1{%
 \Point=\NumStack
 \advance\Point by #1
 \global\ItemNUMBER=\count\Point
}

\def\GetItemTYPE#1{%
 \Point=\TypeStack
 \advance\Point by #1
 \global\ItemTYPE=\count\Point
}

\def\GetItemSPAN#1{%
 \Point\SpanStack
 \advance\Point by #1
 \global\ItemSPAN=\count\Point
}

\def\GetItemBOX#1{%
 \Point=\BoxStack
 \advance\Point by #1
 \global\setbox\ItemBOX=\vbox{\copy\Point}
 \global\ItemSIZE=\ht\ItemBOX
 \global\advance\ItemSIZE by \dp\ItemBOX
 \TEMPCOUNT=\ItemSIZE
 \divide\TEMPCOUNT by \Leading
 \divide\TEMPCOUNT by 65536
 \advance\TEMPCOUNT \@ne
 \ItemSIZE=\TEMPCOUNT pt
 \global\multiply\ItemSIZE by \Leading
}


\def\JoinStack{%
 \ifnum\LengthOfStack=\MaxItems 
  \Warn{WARNING: Stack is full...some items will be lost!}
 \else
  \Point=\StatusStack
  \advance\Point by \LengthOfStack
  \global\count\Point=\ItemSTATUS
  \Point=\NumStack
  \advance\Point by \LengthOfStack
  \global\count\Point=\ItemNUMBER
  \Point=\TypeStack
  \advance\Point by \LengthOfStack
  \global\count\Point=\ItemTYPE
  \Point\SpanStack
  \advance\Point by \LengthOfStack
  \global\count\Point=\ItemSPAN
  \Point=\BoxStack
  \advance\Point by \LengthOfStack
  \global\setbox\Point=\vbox{\copy\ItemBOX}
  \global\advance\LengthOfStack \@ne
  \ifnum\ItemTYPE=\Figure 
   \global\MoreFigurestrue
  \else
   \global\MoreTablestrue
  \fi
 \fi
}


\def\LeaveStack#1{%
 {\Iteration=#1
 \loop
 \ifnum\Iteration<\LengthOfStack
  \advance\Iteration \@ne
  \GetItemSTATUS{\Iteration}
   \advance\Point by \m@ne
   \global\count\Point=\ItemSTATUS
  \GetItemNUMBER{\Iteration}
   \advance\Point by \m@ne
   \global\count\Point=\ItemNUMBER
  \GetItemTYPE{\Iteration}
   \advance\Point by \m@ne
   \global\count\Point=\ItemTYPE
  \GetItemSPAN{\Iteration}
   \advance\Point by \m@ne
   \global\count\Point=\ItemSPAN
  \GetItemBOX{\Iteration}
   \advance\Point by \m@ne
   \global\setbox\Point=\vbox{\copy\ItemBOX}
 \repeat}
 \global\advance\LengthOfStack by \m@ne
}


\newif\ifStackNotClean

\def\CleanStack{%
 \StackNotCleantrue
 {\Iteration=\z@
  \loop
   \ifStackNotClean
    \GetItemSTATUS{\Iteration}
    \ifnum\ItemSTATUS=\InStack
     \advance\Iteration \@ne
     \else
      \LeaveStack{\Iteration}
    \fi
   \ifnum\LengthOfStack<\Iteration
    \StackNotCleanfalse
   \fi
 \repeat}
}


\def\FindItem#1#2{%
 \global\StackPointer=\m@ne 
 {\Iteration=\z@
  \loop
  \ifnum\Iteration<\LengthOfStack
   \GetItemSTATUS{\Iteration}
   \ifnum\ItemSTATUS=\InStack
    \GetItemTYPE{\Iteration}
    \ifnum\ItemTYPE=#1
     \GetItemNUMBER{\Iteration}
     \ifnum\ItemNUMBER=#2
      \global\StackPointer=\Iteration
      \Iteration=\LengthOfStack 
     \fi
    \fi
   \fi
  \advance\Iteration \@ne
 \repeat}
}


\def\FindNext{%
 \global\StackPointer=\m@ne 
 {\Iteration=\z@
  \loop
  \ifnum\Iteration<\LengthOfStack
   \GetItemSTATUS{\Iteration}
   \ifnum\ItemSTATUS=\InStack
    \GetItemTYPE{\Iteration}
   \ifnum\ItemTYPE=\Figure
    \ifMoreFigures
      \global\NextItem=\Figure
      \global\StackPointer=\Iteration
      \Iteration=\LengthOfStack 
    \fi
   \fi
   \ifnum\ItemTYPE=\Table
    \ifMoreTables
      \global\NextItem=\Table
      \global\StackPointer=\Iteration
      \Iteration=\LengthOfStack 
    \fi
   \fi
  \fi
  \advance\Iteration \@ne
 \repeat}
}


\def\ChangeStatus#1#2{%
 \Point=\StatusStack
 \advance\Point by #1
 \global\count\Point=#2
}



\def\Zone{\InZoneA}

\ZoneBAdjust=\z@

\def\MakePage{
 \global\ZoneBSize=\PageHeight
 \global\TextSize=\ZoneBSize
 \global\ZoneAFullPagefalse
 \global\topskip=\TextLeading
 \MakePageInCompletetrue
 \MoreFigurestrue
 \MoreTablestrue
 \FigInZoneBfalse
 \FigInZoneCfalse
 \TabInZoneBfalse
 \TabInZoneCfalse
 \global\FirstSingleItemtrue
 \global\FirstZoneAtrue
 \global\setbox\ZoneABOX=\box\VOIDBOX
 \global\setbox\ZoneBBOX=\box\VOIDBOX
 \global\setbox\ZoneCBOX=\box\VOIDBOX
 \loop
  \ifMakePageInComplete
 \FindNext
 \ifnum\StackPointer=\m@ne
  \NextItem=\m@ne
  \MoreFiguresfalse
  \MoreTablesfalse
 \fi
 \ifnum\NextItem=\Figure
   \FindItem{\Figure}{\NextFigure}
   \ifnum\StackPointer=\m@ne \global\MoreFiguresfalse
   \else
    \GetItemSPAN{\StackPointer}
    \ifnum\ItemSPAN=\Single \def\Zone{\InZoneB}\relax
     \ifFigInZoneC \global\MoreFiguresfalse\fi
    \else
     \def\Zone{\InZoneA}
     \ifFigInZoneB \def\Zone{\InZoneC}\fi
    \fi
   \fi
   \ifMoreFigures\Print{}\FigureItems\fi
 \fi
\ifnum\NextItem=\Table
   \FindItem{\Table}{\NextTable}
   \ifnum\StackPointer=\m@ne \global\MoreTablesfalse
   \else
    \GetItemSPAN{\StackPointer}
    \ifnum\ItemSPAN=\Single\relax
     \ifTabInZoneC \global\MoreTablesfalse\fi
    \else
     \def\Zone{\InZoneA}
     \ifTabInZoneB \def\Zone{\InZoneC}\fi
    \fi
   \fi
   \ifMoreTables\Print{}\TableItems\fi
 \fi
   \MakePageInCompletefalse 
   \ifMoreFigures\MakePageInCompletetrue\fi
   \ifMoreTables\MakePageInCompletetrue\fi
 \repeat
 \ifZoneAFullPage
  \global\TextSize=\z@
  \global\ZoneBSize=\z@
  \global\vsize=\z@\relax
  \global\topskip=\z@\relax
  \vbox to \z@{\vss}
  \eject
 \else
 \global\advance\ZoneBSize by -\ZoneBAdjust
 \global\vsize=\ZoneBSize
 \global\hsize=\ColumnWidth
 \global\ZoneBAdjust=\z@
 \ifdim\TextSize<23pt
 \Warn{}
 \Warn{* Making column fall short: TextSize=\the\TextSize *}
 \vskip-\lastskip\eject\fi
 \fi
}

\def\MakeRightCol{
 \global\TextSize=\ZoneBSize
 \MakePageInCompletetrue
 \MoreFigurestrue
 \MoreTablestrue
 \global\FirstSingleItemtrue
 \global\setbox\ZoneBBOX=\box\VOIDBOX
 \def\Zone{\InZoneB}
 \loop
  \ifMakePageInComplete
 \FindNext
 \ifnum\StackPointer=\m@ne
  \NextItem=\m@ne
  \MoreFiguresfalse
  \MoreTablesfalse
 \fi
 \ifnum\NextItem=\Figure
   \FindItem{\Figure}{\NextFigure}
   \ifnum\StackPointer=\m@ne \MoreFiguresfalse
   \else
    \GetItemSPAN{\StackPointer}
    \ifnum\ItemSPAN=\Double\relax
     \MoreFiguresfalse\fi
   \fi
   \ifMoreFigures\Print{}\FigureItems\fi
 \fi
 \ifnum\NextItem=\Table
   \FindItem{\Table}{\NextTable}
   \ifnum\StackPointer=\m@ne \MoreTablesfalse
   \else
    \GetItemSPAN{\StackPointer}
    \ifnum\ItemSPAN=\Double\relax
     \MoreTablesfalse\fi
   \fi
   \ifMoreTables\Print{}\TableItems\fi
 \fi
   \MakePageInCompletefalse 
   \ifMoreFigures\MakePageInCompletetrue\fi
   \ifMoreTables\MakePageInCompletetrue\fi
 \repeat
 \ifZoneAFullPage
  \global\TextSize=\z@
  \global\ZoneBSize=\z@
  \global\vsize=\z@\relax
  \global\topskip=\z@\relax
  \vbox to \z@{\vss}
  \eject
 \else
 \global\vsize=\ZoneBSize
 \global\hsize=\ColumnWidth
 \ifdim\TextSize<23pt
 \Warn{}
 \Warn{* Making column fall short: TextSize=\the\TextSize *}
 \vskip-\lastskip\eject\fi
\fi
}

\def\FigureItems{
 \Print{Considering...}
 \ShowItem{\StackPointer}
 \GetItemBOX{\StackPointer} 
 \GetItemSPAN{\StackPointer}
  \CheckFitInZone 
  \ifnum\ItemFits=\Yes
   \ifnum\ItemSPAN=\Single
     \ChangeStatus{\StackPointer}{\InZoneB} 
     \global\FigInZoneBtrue
     \ifFirstSingleItem
      \hbox{}\vskip-\BodgeHeight
     \global\advance\ItemSIZE by \TextLeading
     \fi
     \unvbox\ItemBOX\ItemSep
     \global\FirstSingleItemfalse
     \global\advance\TextSize by -\ItemSIZE
     \global\advance\TextSize by -\TextLeading
   \else
    \ifFirstZoneA
     \global\advance\ItemSIZE by \TextLeading
     \global\FirstZoneAfalse\fi
    \global\advance\TextSize by -\ItemSIZE
    \global\advance\TextSize by -\TextLeading
    \global\advance\ZoneBSize by -\ItemSIZE
    \global\advance\ZoneBSize by -\TextLeading
    \ifFigInZoneB\relax
     \else
     \ifdim\TextSize<3\TextLeading
     \global\ZoneAFullPagetrue
     \fi
    \fi
    \ChangeStatus{\StackPointer}{\Zone}
    \ifnum\Zone=\InZoneC \global\FigInZoneCtrue\fi
  \fi
   \Print{TextSize=\the\TextSize}
   \Print{ZoneBSize=\the\ZoneBSize}
  \global\advance\NextFigure \@ne
   \Print{This figure has been placed.}
  \else
   \Print{No space available for this figure...holding over.}
   \Print{}
   \global\MoreFiguresfalse
  \fi
}

\def\TableItems{
 \Print{Considering...}
 \ShowItem{\StackPointer}
 \GetItemBOX{\StackPointer} 
 \GetItemSPAN{\StackPointer}
  \CheckFitInZone 
  \ifnum\ItemFits=\Yes
   \ifnum\ItemSPAN=\Single
    \ChangeStatus{\StackPointer}{\InZoneB}
     \global\TabInZoneBtrue
     \ifFirstSingleItem
      \hbox{}\vskip-\BodgeHeight
     \global\advance\ItemSIZE by \TextLeading
     \fi
     \unvbox\ItemBOX\ItemSep
     \global\FirstSingleItemfalse
     \global\advance\TextSize by -\ItemSIZE
     \global\advance\TextSize by -\TextLeading
   \else
    \ifFirstZoneA
    \global\advance\ItemSIZE by \TextLeading
    \global\FirstZoneAfalse\fi
    \global\advance\TextSize by -\ItemSIZE
    \global\advance\TextSize by -\TextLeading
    \global\advance\ZoneBSize by -\ItemSIZE
    \global\advance\ZoneBSize by -\TextLeading
    \ifFigInZoneB\relax
     \else
     \ifdim\TextSize<3\TextLeading
     \global\ZoneAFullPagetrue
     \fi
    \fi
    \ChangeStatus{\StackPointer}{\Zone}
    \ifnum\Zone=\InZoneC \global\TabInZoneCtrue\fi
   \fi
  \global\advance\NextTable \@ne
   \Print{This table has been placed.}
  \else
  \Print{No space available for this table...holding over.}
   \Print{}
   \global\MoreTablesfalse
  \fi
}


\def\CheckFitInZone{%
{\advance\TextSize by -\ItemSIZE
 \advance\TextSize by -\TextLeading
 \ifFirstSingleItem
  \advance\TextSize by \TextLeading
 \fi
 \ifnum\Zone=\InZoneA\relax
  \else \advance\TextSize by -\ZoneBAdjust
 \fi
 \ifdim\TextSize<3\TextLeading \global\ItemFits=\No
 \else \global\ItemFits=\Yes\fi}
}

\def\BeginOpening{%
  \thispagestyle{titlepage}%
  \global\setbox\ItemBOX=\vbox\bgroup%
    \hsize=\PageWidth%
    \hrule height \z@
    \ifsinglecol\vskip 6pt\fi 
}

\let\begintopmatter=\BeginOpening  

\def\EndOpening{%
  \One
  \egroup
  \ifsinglecol
    \box\ItemBOX%
    \vskip\TextLeading plus 2\TextLeading
    \@noafterindent
  \else
    \ItemNUMBER=\z@%
    \ItemTYPE=\Figure
    \ItemSPAN=\Double
    \ItemSTATUS=\InStack
    \JoinStack
  \fi
}


\newif\if@here  \@herefalse

\def\no@float{\global\@heretrue}
\let\nofloat=\relax 

\def\beginfigure{%
  \@ifstar{\global\@dfloattrue \@bfigure}{\global\@dfloatfalse \@bfigure}%
}

\def\@bfigure#1{%
  \par
  \if@dfloat
    \ItemSPAN=\Double
    \TEMPDIMEN=\PageWidth
  \else
    \ItemSPAN=\Single
    \TEMPDIMEN=\ColumnWidth
  \fi
  \ifsinglecol
    \TEMPDIMEN=\PageWidth
  \else
    \ItemSTATUS=\InStack
    \ItemNUMBER=#1%
    \ItemTYPE=\Figure
  \fi
  \bgroup
    \hsize=\TEMPDIMEN
    \global\setbox\ItemBOX=\vbox\bgroup
      \eightpoint\nostb@ls{10pt}%
      \let\caption=\fig@caption
      \ifsinglecol \let\nofloat=\no@float\fi
}

\def\fig@caption#1{%
  \vskip 5.5pt plus 6pt%
  \bgroup 
    \eightpoint\nostb@ls{10pt}%
    \setbox\TEMPBOX=\hbox{#1}%
    \ifdim\wd\TEMPBOX>\TEMPDIMEN
      \noindent \unhbox\TEMPBOX\par
    \else
      \hbox to \hsize{\hfil\unhbox\TEMPBOX\hfil}%
    \fi
  \egroup
}

\def\endfigure{%
  \par\egroup 
  \egroup
  \ifsinglecol
    \if@here \midinsert\global\@herefalse\else \topinsert\fi
      \unvbox\ItemBOX
    \endinsert
  \else
    \JoinStack
    \Print{Processing source for figure \the\ItemNUMBER}%
  \fi
}


\newbox\tab@cap@box
\def\tab@caption#1{\global\setbox\tab@cap@box=\hbox{#1\par}}

\newtoks\tab@txt@toks
\long\def\tab@txt#1{\global\tab@txt@toks={#1}\global\table@txttrue}

\newif\iftable@txt  \table@txtfalse
\newif\if@dfloat    \@dfloatfalse

\def\begintable{%
  \@ifstar{\global\@dfloattrue \@btable}{\global\@dfloatfalse \@btable}%
}

\def\@btable#1{%
  \par
  \if@dfloat
    \ItemSPAN=\Double
    \TEMPDIMEN=\PageWidth
  \else
    \ItemSPAN=\Single
    \TEMPDIMEN=\ColumnWidth
  \fi
  \ifsinglecol
    \TEMPDIMEN=\PageWidth
  \else
    \ItemSTATUS=\InStack
    \ItemNUMBER=#1%
    \ItemTYPE=\Table
  \fi
  \bgroup
    \eightpoint\nostb@ls{10pt}%
    \global\setbox\ItemBOX=\vbox\bgroup
      \let\caption=\tab@caption
      \let\tabletext=\tab@txt
      \ifsinglecol \let\nofloat=\no@float\fi
}

\def\endtable{%
  \par\egroup 
  \egroup
  \setbox\TEMPBOX=\hbox to \TEMPDIMEN{%
    \hss
    \vbox{%
      \hsize=\wd\ItemBOX
      \ifvoid\tab@cap@box
      \else
        \noindent\unhbox\tab@cap@box
        \vskip 5.5pt plus 6pt%
      \fi
      \box\ItemBOX
      \iftable@txt
        \vskip 10pt%
        \eightpoint\nostb@ls{10pt}%
        \noindent\the\tab@txt@toks
        \global\table@txtfalse
      \fi
    }%
    \hss
  }%
  \ifsinglecol
    \if@here \midinsert\global\@herefalse\else \topinsert\fi
      \box\TEMPBOX
    \endinsert
  \else
    \global\setbox\ItemBOX=\box\TEMPBOX
    \JoinStack
    \Print{Processing source for table \the\ItemNUMBER}%
  \fi
}

\def\UnloadZoneA{%
\FirstZoneAtrue
 \Iteration=\z@
  \loop
   \ifnum\Iteration<\LengthOfStack
    \GetItemSTATUS{\Iteration}
    \ifnum\ItemSTATUS=\InZoneA
     \GetItemBOX{\Iteration}
     \ifFirstZoneA \vbox to \BodgeHeight{\vfil}%
     \FirstZoneAfalse\fi
     \unvbox\ItemBOX\ItemSep
     \LeaveStack{\Iteration}
     \else
     \advance\Iteration \@ne
   \fi
 \repeat
}

\def\UnloadZoneC{%
\Iteration=\z@
  \loop
   \ifnum\Iteration<\LengthOfStack
    \GetItemSTATUS{\Iteration}
    \ifnum\ItemSTATUS=\InZoneC
     \GetItemBOX{\Iteration}
     \ItemSep\unvbox\ItemBOX
     \LeaveStack{\Iteration}
     \else
     \advance\Iteration \@ne
   \fi
 \repeat
}


\def\ShowItem#1{
  {\GetItemAll{#1}
  \Print{\the#1:
  {TYPE=\ifnum\ItemTYPE=\Figure Figure\else Table\fi}
  {NUMBER=\the\ItemNUMBER}
  {SPAN=\ifnum\ItemSPAN=\Single Single\else Double\fi}
  {SIZE=\the\ItemSIZE}}}
}

\def\ShowStack{%
 \Print{}
 \Print{LengthOfStack = \the\LengthOfStack}
 \ifnum\LengthOfStack=\z@ \Print{Stack is empty}\fi
 \Iteration=\z@
 \loop
 \ifnum\Iteration<\LengthOfStack
  \ShowItem{\Iteration}
  \advance\Iteration \@ne
 \repeat
}

\def\B#1#2{%
\hbox{\vrule\kern-0.4pt\vbox to #2{%
\hrule width #1\vfill\hrule}\kern-0.4pt\vrule}
}


\newif\ifsinglecol   \singlecolfalse

\def\onecolumn{%
  \global\output={\singlecoloutput}%
  \global\hsize=\PageWidth
  \global\vsize=\PageHeight
  \global\ColumnWidth=\hsize
  \global\TextLeading=12pt
  \global\Leading=12
  \global\singlecoltrue
  \global\let\onecolumn=\relax
  \global\let\footnote=\sing@footnote
  \global\let\vfootnote=\sing@vfootnote
  \ninepoint 
  \message{(Single column)}%
}

\def\singlecoloutput{%
  \shipout\vbox{\PageHead\pagebody\PageFoot}%
  \advancepageno
  \ifplate@page
    \shipout\vbox{%
      \sp@pagetrue
      \def\sp@type{plate}%
      \global\plate@pagefalse
      \PageHead\vbox to \PageHeight{\unvbox\plt@box\vfil}\PageFoot%
    }%
    \message{[plate]}%
    \advancepageno
  \fi
  \ifnum\outputpenalty>-\@MM \else\dosupereject\fi%
}

\def\ItemSep{\vskip\ItemSepamount\relax}

\def\ItemSepbreak{\par\ifdim\lastskip<\ItemSepamount
  \removelastskip\penalty-200\ItemSep\fi%
}


\let\@@endinsert=\endinsert 

\def\endinsert{\egroup 
  \if@mid \dimen@\ht\z@ \advance\dimen@\dp\z@ \advance\dimen@12\p@
    \advance\dimen@\pagetotal \advance\dimen@-\pageshrink
    \ifdim\dimen@>\pagegoal\@midfalse\p@gefalse\fi\fi
  \if@mid \ItemSep\box\z@\ItemSepbreak
  \else\insert\topins{\penalty100 
    \splittopskip\z@skip
    \splitmaxdepth\maxdimen \floatingpenalty\z@
    \ifp@ge \dimen@\dp\z@
    \vbox to\vsize{\unvbox\z@\kern-\dimen@}
    \else \box\z@\nobreak\ItemSep\fi}\fi\endgroup%
}


\def\gobbleone#1{}
\def\gobbletwo#1#2{}
\let\footnote=\gobbletwo 
\let\vfootnote=\gobbleone

\def\sing@footnote#1{\let\@sf\empty 
  \ifhmode\edef\@sf{\spacefactor\the\spacefactor}\/\fi
  \hbox{$^{\hbox{\eightpoint #1}}$}\@sf\sing@vfootnote{#1}%
}

\def\sing@vfootnote#1{\insert\footins\bgroup\eightpoint\b@ls{9pt}%
  \interlinepenalty\interfootnotelinepenalty
  \splittopskip\ht\strutbox 
  \splitmaxdepth\dp\strutbox \floatingpenalty\@MM
  \leftskip\z@skip \rightskip\z@skip \spaceskip\z@skip \xspaceskip\z@skip
  \noindent $^{\scriptstyle\hbox{#1}}$\hskip 4pt%
    \footstrut\futurelet\next\fo@t%
}

\def\footnoterule{\kern-3\p@ \hrule height \z@ \kern 3\p@}

\skip\footins=19.5pt plus 12pt minus 1pt
\count\footins=1000
\dimen\footins=\maxdimen


\def\landscape{%
  \global\TEMPDIMEN=\PageWidth
  \global\PageWidth=\PageHeight
  \global\PageHeight=\TEMPDIMEN
  \global\let\landscape=\relax
  \onecolumn
  \message{(landscape)}%
  \raggedbottom
}


\output{%
  \ifLeftCOL
    \global\setbox\LeftBOX=\vbox to \ZoneBSize{\box255\unvbox\ZoneBBOX}%
    \global\LeftCOLfalse
    \MakeRightCol
  \else
    \setbox\RightBOX=\vbox to \ZoneBSize{\box255\unvbox\ZoneBBOX}%
    \setbox\MidBOX=\hbox{\box\LeftBOX\hskip\ColumnGap\box\RightBOX}%
    \setbox\PageBOX=\vbox to \PageHeight{%
      \UnloadZoneA\box\MidBOX\UnloadZoneC}%
    \shipout\vbox{\PageHead\box\PageBOX\PageFoot}%
    \advancepageno
    \ifplate@page
      \shipout\vbox{%
        \sp@pagetrue
        \def\sp@type{plate}%
        \global\plate@pagefalse
        \PageHead\vbox to \PageHeight{\unvbox\plt@box\vfil}\PageFoot%
      }%
      \message{[plate]}%
      \advancepageno
    \fi
    \global\LeftCOLtrue
    \CleanStack
    \MakePage
  \fi
}


\Warn{\start@mess}


\catcode `\@=12 



\hoffset=-.5cm
\voffset=.5cm
\font\fivebmi=cmmib6
\font\sixbmi=cmmib6	\skewchar\sixbmi='177
\font\ninebmi=cmmib10 at 9pt 	\skewchar\ninebmi='177
\newfam\bmifam
\textfont\bmifam=\ninebmi
\scriptfont\bmifam=\sixbmi
\scriptscriptfont\bmifam=\fivebmi
\def\bmi{\fam\bmifam\ninebmi}
\def\b#1{{\bmi#1}}

\mathchardef\alpha="710B
\mathchardef\beta="710C
\mathchardef\gamma="710D
\mathchardef\delta="710E
\mathchardef\epsilon="710F
\mathchardef\zeta="7110
\mathchardef\eta="7111
\mathchardef\theta="7112
\mathchardef\iota="7113
\mathchardef\kappa="7114
\mathchardef\lambda="7115
\mathchardef\mu="7116
\mathchardef\nu="7117
\mathchardef\xi="7118
\mathchardef\pi="7119
\mathchardef\rho="711A
\mathchardef\sigma="711B
\mathchardef\tau="711C
\mathchardef\upsilon="711D
\mathchardef\phi="711E
\mathchardef\chi="711F
\mathchardef\psi="7120
\mathchardef\omega="7121
\mathchardef\varepsilon="7122
\mathchardef\vartheta="7123
\mathchardef\varpi="7124
\mathchardef\varrho="7125
\mathchardef\varsigma="7126
\mathchardef\varphi="7127

\def\chaphead{}
\newcount\eqnumber
\eqnumber=1

\def\today{\ifcase\month\or
 January\or February\or March\or April\or May\or June\or
 July\or August\or September\or October\or November\or December\fi
 \space\number\day, \number\year}

\def\eqnam#1{\xdef#1{(\chaphead\the\eqnumber}}

\def\newe{(\hbox{\chaphead\the\eqnumber})\global\advance\eqnumber by 1}
\def\firste{(\hbox{\chaphead\the\eqnumber a})\global\advance\eqnumber by 1}
\def\laste#1{\advance\eqnumber by -1%
	(\hbox{\chaphead\the\eqnumber #1})\advance\eqnumber by 1}

\def\refe#1{\advance\eqnumber by -#1 (\chaphead\the\eqnumber
     \advance\eqnumber by #1 }

\def\disp{\displaystyle}

\def\i{\relax\ifmmode{\rm i}\else\char16\fi}
\def\e{{\rm e}}
\def\frac#1#2{{\textstyle{#1\over#2}}}

\def\d{{\rm d}}
\def\dddot#1{\ddot#1\kern-1.4pt\dot{\phantom{#1}}\kern-3pt}

\def\erf{\mathop{\rm erf}\nolimits} 


\def\spose#1{\hbox to 0pt{#1\hss}}

\def\=#1{\overline{#1}}

\def\lta{\mathrel{\spose{\lower 3pt\hbox{$\mathchar"218$}}
     \raise 2.0pt\hbox{$\mathchar"13C$}}}
\def\gta{\mathrel{\spose{\lower 3pt\hbox{$\mathchar"218$}}
     \raise 2.0pt\hbox{$\mathchar"13E$}}}

\def\kms{{\rm\,km\,s^{-1}}}
\def\kpc{{\rm\,kpc}}

\def\msun{{\rm\,M_\odot}}

\def\Gyr{{\rm\,Gyr}}

\def\annrev #1 #2 {ARA\&A, #1, #2}
\def\aa #1 #2 {A\&A, #1, #2}
\def\aasupp #1 #2 {A\&AS, #1, #2}
\def\aj #1 #2 {AJ, #1, #2}
\def\apj #1 #2 {ApJ, #1, #2}
\def\apjlett #1 #2 {ApJ, #1, #2}
\def\apjsupp #1 #2 {ApJS, #1, #2}
\def\ban #1 #2 {Bull.\ Astron.\ Inst.\ Netherlands, #1, #2}
\def\mn #1 #2 {MNRAS, #1, #2}
\def\nature #1 #2 {Nat, #1, #2}
\def\pasj #1 #2 {PASJ, #1, #2}
\def\pasp #1 #2 {PASP, #1, #2}

\input psfig

\overfullrule=0pt
\newif\ifpsfiles\psfilestrue

\def\frac#1#2{{#1\over#2}}
\def\label#1{}\def\cite#1{#1}

\def\c{{\rm c}}\def\d{{\rm d}}
\def\t{{\rm t}}

\begintopmatter
\title{The Orbit and Mass of the Sagittarius Dwarf Galaxy}

\author{Ing-Guey Jiang and James Binney}
\affiliation{Theoretical Physics, University of Oxford, Oxford, OX1 3NP} 
\shortauthor{I.-G.\ Jiang and J.J.\ Binney}
\shorttitle{Sagittarius dwarf galaxy}

\abstract{Possible orbital histories of the Sgr dwarf galaxy are explored.
A special-purpose $N$-body code is used to construct the first models of the
Milky Way -- Sgr Dwarf system in which both the Milky Way and the Sgr Dwarf
are represented by full $N$-body systems and followed for a Hubble time.
These models are used to calibrate a semi-analytic model of the Dwarf's
orbit that enable us to explore a wider parameter space than is accessible
to the $N$-body models. We conclude that the extant data on the Dwarf are
compatible with a wide range of orbital histories. At one extreme the Dwarf
initially possesses $\sim10^{11}\msun$ and starts from a Galactocentric
distance $R_{\rm D}(0)\gta200\kpc$. At the other extreme the Dwarf starts
with $\sim10^9\msun$ and $R_{\rm D}(0)\sim60\kpc$, similar to its present
apocentric distance. In all cases the Dwarf is initially dark-matter
dominated and the current velocity dispersion of the Dwarf's dark matter is
tightly constrained to be $(21\pm2)\kms$.  This number is probably
compatible with the smaller measured dispersion of the Dwarf's stars because
of (a) the dynamical difference between dark and luminous matter, and (b)
velocity anisotropy.}

\keywords{Galaxy: kinematics and dynamics -- Galaxy: halo 
-- galaxies: individual: Sgr Dwarf -- galaxies: dwarf -- galaxies: formation}

\maketitle

\section{Introduction}

The Milky Way's nearest neighbour, the Sagittarius dwarf galaxy, lies only
$16\kpc$ from the Galactic centre, but was until 1993 hidden from us by the
inner Milky Way. Since the discovery of this object by Ibata et al.\ (1994),
several studies have explored its extent on the sky, its mean radial
velocity, its proper motion perpendicular to the Galactic plane, and its
internal velocity dispersion (Ibata et al.\ 1997). The data currently in
hand constrain the present orbit of the Dwarf quite tightly.  Ibata \& Lewis
(1998) find that acceptable orbits have periods $\lta1\Gyr$
and are moderately eccentric, with apocentres near $60\kpc$ and pericentres
near $20\kpc$.

Moving on its orbit the Dwarf is subject to significant tidal distortion by
the Milky Way. Indeed, the observed elongation of the Dwarf perpendicular to
the Galactic plane is thought to be the result of tidal shear. In this
context it is important to understand how the Dwarf has avoided being torn
apart by Galactic tides. Vel\'asquez \& White (1995), Johnson et al.~(1995)
and Ibata \& Lewis (1998) have studied this problem and concluded that
there is at most a tight corner of parameter space in which the Dwarf could
have survived to the present time. Specifically, if light is assumed to
trace mass, survival is impossible: to ensure survival it is essential to
pack as much mass as possible within the observed outer radius, $r_\t$, of
the Dwarf.

The observed internal velocity dispersion places an upper limit on the
Dwarf's central mass density, so to maximize the Dwarf's mass one has to
pack the material around Dwarf's edge. Hence the extra mass must be dark and
within the Dwarf's outer limit its density must decrease outwards as slowly
as possible. Since material beyond the tidal radius is not bound to the
Dwarf, the density of dark matter should plummet near $r_\t$. Ibata \& Lewis
found a density distribution that was consistent with these requirements and
has a non-negative distribution function $f(E)$:
$\rho(r)\propto(\e^{-(r/1\kpc)^2}-\e^{-1})$. They present $N$-body
simulations of this dark-matter distribution moving on the Dwarf's orbit,
and show that it is torn apart by the Galaxy on an acceptably long timescale.

The model of Ibata \& Lewis is attractive because we know that dark matter
contributes significantly to the potentials of dwarf galaxies.  It is,
however, finely tuned in that both the density and the radial extent of the
dark-matter distribution can be neither larger nor smaller than the chosen
values. This fine tuning would not detract from the plausibility of the
model if it arose naturally as the Dwarf's orbit and the density profile
were fashioned by Galactic tides and dynamical friction against the Galactic
halo. In this paper we present simulations designed to investigate this
question.

Specifically, we aim to find initial configurations in which the Dwarf's
halo encompasses a conventional flat-rotation-curve section in addition to a
homogeneous core. We wish to simulate the stripping of this halo to leave
the sharp-edged homogeneous rump envisaged by Ibata \& Lewis.  Full $N$-body
simulations of this process are extremely costly because the Galaxy's dark
halo has to be simulated out to a Galactocentric radius $r\gta250\kpc$,
within which its mass is $\sim 2\times10^{12}\msun$, while simultaneously
following the internal dynamics of the Dwarf, which now contains $\lta
10^8\msun $ of visible material and probably a comparable amount of dark
matter. Resolution of the visible Dwarf into $\gta100$ particles, implies
that individual particles have masses $\sim 10^6\msun$, so $\gta2\times10^6$
such particles are required to represent the Galactic halo. This large
number of particles has to be followed for a Hubble time, or $\sim400$
current half-light crossing times of the Dwarf. In principle one may
represent the Galactic halo by fewer particles of higher mass, but this
strategy is unsafe because in such a simulation two-particle relaxation
proceeds rapidly and artificially accelerates the disruption of the Dwarf,
which is a process of prime interest.

Previous simulations of the Dwarf's orbit have relied on a number of
more-or-less unsatisfactory work-arounds, such as treating the Galactic
potential as fixed, and possibly augmented by dynamical friction (Vel\'asquez
\& White 1995, Johnston et al.\ 1995, Ibata \& Lewis 1998), studying only
orbits that never reach far out into the Galactic halo and using more
massive particles for the halo than for the Dwarf (G\'omez-Flechoso, Fux \&
Martinet, 1999). Our strategy has been two-fold. First we have tailored a
potential solver to the problem: this uses two multipole expansions, one
centred on the Galactic centre and one centred on the Dwarf. Second we have
used a small number of large $N$-body simulations to calibrate
semi-analytic calculations that include dynamical friction and tidal
stripping, and have used the semi-analytic calculations to explore more
thoroughly the parameter space associated with the initial Dwarf and its
orbit.

The paper is organized as follows. Section 2 summarizes the mass profiles
assumed for the Milky Way and the Dwarf. Section 3 describes the $N$-body
code that was tailored to the problem, together with tests of the code
and three full $N$-body models of the Galaxy--Dwarf system.  Section 4 
describes a semi-analytic model calibrated by reference to the $N$-body
models. Section 5 sums up and discusses the observable consequences of
extensive mass loss by the Dwarf.

\section{Initial Conditions}

To be possessed of an extensive halo, the Dwarf must initially be at a much
larger Galactocentric radius than at present. Hence the orbit must initially
have had a large apocentre, and therefore have been a long-period orbit.
Since we know that the Dwarf is now on a short-period orbit, it must have
lost significant orbital energy. Zhao (1998) suggested that a close
encounter with the Magellanic Clouds could have led to this loss of orbital
energy.  This possibility seems unlikely, however, because the gravitational
field of the Clouds is probably too weak to deflect the Dwarf through a
significant angle, given the relative velocity ($\sim300\kms$) at which the
Dwarf would have encountered the Clouds -- see the simulations of Ibata \&
Lewis for support of this view. Given our belief that the Dwarf will
initially have possessed an extensive dark halo, the natural mechanism for
loss of orbital energy is dynamical friction.

If the dark-halo model of Ibata \& Lewis is correct, the current mass of the
Dwarf is $\sim10^9\msun$. Dynamical friction against the Galactic halo has
only a modest effect on a body of this mass. For example, by equation (7-27)
of Binney \& Tremaine (1987; hereafter BT), its decay time from a circular
orbit of initial radius $30\kpc$ is $(50/\ln\Lambda)\Gyr$ with
$\ln\Lambda\sim8$ (see below), and increases as the square of the initial
radius.  Hence, if the mass of the Dwarf were initially as small as it now
probably is, its orbit would not have evolved very much, and it could never
have possessed a generic dark halo.

If, by contrast, the Dwarf started out much more massive, its orbit could
have evolved from the large galactocentric radius. At such a large radius it
could have possessed the extensive halo that alone would make it massive.
Hence, it is a priori plausible that there are a number of self-consistent
solutions for the Dwarf's past: at one extreme it was from the outset
severely tidally truncated and has moved at all times on the same
short-period orbit; at the other extreme, it was initially possessed of a
massive power-law halo that caused it to sink rapidly inwards under the
influence of dynamical friction, and be progressively stripped of its halo.
Ibata \& Lewis have presented an orbit of the first kind. We seek 
orbits of the second kind.

\subsection{Initial density profiles and velocities}

The Galactic disk is assumed to be
rigid. The surface density of the disk is taken to be exponential inside
$R_\t=19.75\kpc$ and then to fall rapidly to zero by $R_0=20\kpc$.
In terms of the variable
$$
x\equiv{\pi\over2}{R-R_r\over R_0-R_\t},
\eqno\firste$$
the disk's surface density is given by
$$
\Sigma(R)=\Sigma_0\e^{-R/R_\d}\times\cases{1&$x\le0$\cr
\cos^2x&$0<x\le\pi/2$\cr
0&otherwise.
}\eqno\laste b$$
We set the disk's scale-length to $R_\d=3.4\kpc$, which yields a
halo-dominated rotation curve (Dehnen \& Binney 1997).  
The parameter $\Sigma_0$ is set such that the disk's total mass is
$10^{11}\msun$.

\begintable1
\caption{{\bf Table 1.} Parameters of the model Galaxy}
\halign to 5.6cm%
{\hfil$#$\ \hfil&\hfil$#$\ \hfil&\hfil$#$\ \hfil&\hfil$#$\hfil&\hfil$#$\hfil\cr
v_\c/\!\kms&r_\c/\!\kpc&r_0/\!\kpc&r_\infty/\!\kpc&M/\!\msun\cr
\noalign{\vskip2pt\hrule\vskip1pt}
181&1&200&1000&2.5\times10^{12}\cr
\noalign{\vskip2pt\hrule}
}
\endtable

Flat rotation curves imply that dark halos have approximately isothermal
density profiles. Therefore we assume that the initial density profiles of
both the Dwarf and the Milky Way are given by
 $$\eqnam\initdprof
\rho(r)=\cases{\disp{v_c^2\over4\pi G}{\e^{-r/r_0}\over r^2+r_c^2}&for
$r\le r_\infty$\cr 
{\disp0\phantom{1\over2}}&otherwise.}
\eqno\newe$$
Table 1 gives values for the Galaxy of the parameters $v_\c$, $r_\c$, $r_0$
and $r_\infty$.  These values are chosen to ensure that the rotation curve
is compatible with observations (Dehnen \& Binney, 1997) and that the mass
of the Local Group is as large as the timing argument implies (e.g.,
Schmoldt \& Saha, 1998). Fig.~1 shows the overall circular-speed curve and
its contributions from disk and halo.

\beginfigure1
\centerline{\psfig{file=rotc.ps,width=\hsize}}
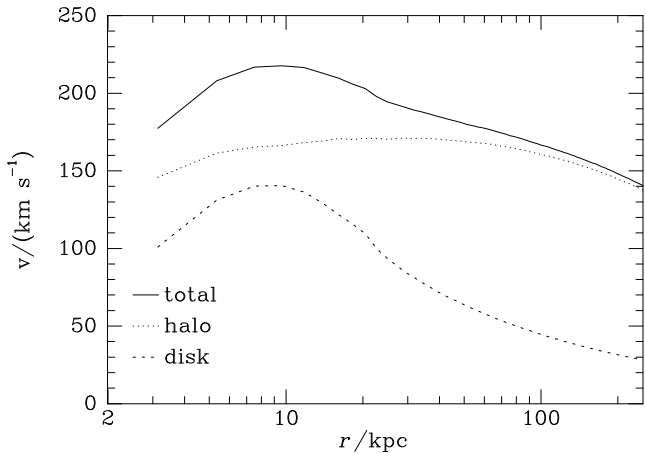
\caption{{\bf Figure 1.} The circular-speed curve of the Milky Way and its
decomposition into contributions from the disc and halo.}
\endfigure

The Dwarf's core radius is taken to be $r_\c=1\kpc$. The other parameters of
the dwarf vary from simulation to simulation and are given in Table 2.

\section{N-body Code}

Our $N$-body code is based on the Poisson solver described by Bontekoe
(1988). Specifically, the particle positions are used to evaluate the
density on a spherical grid. Then the density within each spherical shell is
expanded in spherical harmonics, with harmonics up to $l=8$ included. This
done, the potential at any point can be obtained from BT eq.~(2-122). Our
code differs from that of Bontekoe in the following ways.

\item{1.} Both the angular and the radial grids are adaptive. The radial
grid points move so that roughly equal numbers of particles
lie in each interval of the radial grid. Within each radial bin, a grid in
colatitude $\theta$ is established that has roughly equal
numbers of particles in each bin, and within each of these bins a grid in
azimuth $\phi$ is established that has equal numbers of particles in each
bin.

\item{2.} There are separate grids for particles belonging to the Galaxy and
the Dwarf: the Galaxy's grid has 100 radial points, 24 in
$\theta$ and 48 in $\phi$; the corresponding numbers for the Dwarf's grid
are 12, 12, and 24.
At each time-step, the mass distribution due to Galaxy particles
is determined on the Galaxy grid. The corresponding potential is then
found and used to calculate the forces that the Galaxy imposes on each
Dwarf particle. Then the mass of the Dwarf particles is added to the Galaxy
grid, the potential redetermined and used to calculate the forces experienced
by Galaxy particles. Finally, the mass distribution due to Dwarf particles
is determined on the Dwarf grid, the corresponding potential found and used
to calculate the forces on Dwarf particles from Dwarf particles.

\item{3.} The algorithm of Quinn et al (1997) is used to advance the
particles with timesteps of length $2^n\Delta_{\rm min}$, where
$n=0,1,\ldots,4$. 

As a simulation proceeds, particles are stripped from the Dwarf and one
needs a method of identifying the particles that remain bound to the Dwarf.
The extent, $r_\t$, of the Dwarf is the radius of the largest sphere, centred
the last position of the Dwarf's centre, within which the mean density of all
particles, Dwarf and Galactic, exceeds the mean density of the Galaxy within
the Galactocentric sphere that touches the centre of the Dwarf. The new
position and velocity of the Dwarf are the mean position and velocity of the
Dwarf particles that lie within the Dwarf's sphere.

\beginfigure2
\centerline{\psfig{file=density.ps,width=\hsize}}
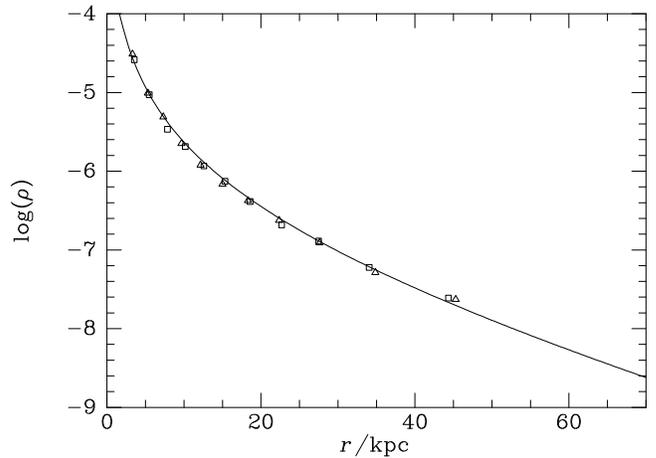
\caption{{\bf Figure 2.} An initial density profile of the Dwarf (a) as an
analytic function (curve), (b) as initially sampled (triangles), and (c) after
$2.1\Gyr$ (squares).}
\endfigure

The simulations represent the Galaxy with $300\,000$ particles and the Dwarf
with $12\,000(M_{\rm D}/10^{11}\msun)$ particles.

\subsection{Tests}

Over $11\Gyr$, which is $170$ times the central free-fall time
$(G\rho)^{-1/2}$ of the Dwarf, the total energy in the simulation changes by
$\lta1.7$ percent.

In the absence of potential softening and tidal limitation, the initial
configuration of the Dwarf would be a self-consistent equilibrium. Fig.~2
shows the degree to which the Dwarf's density profile changes in the first
few dynamical times of a simulation. Specifically: the curve shows the
analytic density profile [eq.~\initdprof)] that we seek to represent; the
triangles show the density profile one infers from the initial positions of
the particles; the squares show the density profile inferred $2.1\Gyr$
later. One sees that any change in the density profile is no larger than the
errors inherent in Monte-Carlo sampling the underlying analytic profile.

\begintable*2
\caption{{\bf Table 2.} Parameters of the orbits. Orbits with upper-case
labels in column 1 have been followed with $N$-body simulations in addition
to the semi-analytic model.}
\halign to 10cm%
{\hfil#\hfil\ &\hfil#\hfil\ &\hfil#\hfil\ &\hfil#\hfil\ &\hfil#\hfil\ &\hfil#\hfil\ &\hfil#\hfil\ &\hfil#\hfil\ &\hfil#\hfil\cr
Orbit&
${\disp{ M_{\rm D\infty}\over10^{10}\msun}}$&
${\disp{ M_{\rm D}(0)\over10^{10}\msun}}$&
${\disp{ r_{\rm D0}\over{\rm kpc}}}$&
${\disp{ r_{{\rm D}\infty}\over{\rm kpc}}}$&
${\disp{ R_{\rm D}(0)\over{\rm kpc}}}$&
${\disp{ t_{\rm sink}\over{\rm Gyr}}}$&
${\disp{ M_{\rm D}(t_{\rm sink})\over10^9\msun}}$&
${\disp{ \sigma_{\rm D}(t_{\rm sink})\over{\rm km\,s^{-1}}}}$\cr
\noalign{\vskip2pt\hrule\vskip2pt}
A&$10$&10\ &20&70&250&11.1&$2.0$     &23.4 \cr
b&$10$&9.8&25&80&250&13.4&$1.1$&18.6  \cr
c&$10$&9.9&20&70&225&\ 8.9&$2.9$   &26.1  \cr
d&$10$&9.7&25&80&225&10.9&$1.7$  &21.7  \cr
e&$10$&9.4&25&80&200&\ 8.6&$1.9$   &22.8  \cr
F&$10$&8.8&30&100&200&10.6&$1.1$ &18.8  \cr
g&$\ 9$ &8.4&25&80&200&\ 9.9&$1.8$   &21.9  \cr
h&$\ 7$ &6.4&25&80&200&14.1&$1.0$  &17.9  \cr
i&$\ 7$ &6.7&20&70&200&10.8&$1.8$  &22.0  \cr
j&$\ 7$ &6.3  &20&70&150&\ 5.9&$3.1$ &27.0  \cr
K&$\ 5$ &4.4  &20&70&150&\ 9.9&$1.8$ &21.9  \cr
l&$\ 5$&3.7&20&70&100&\ 4.5&$3.3$    &27.9  \cr
m&$\ 3$&2.0&20&70&100&10.1&$1.7$   &21.4 \cr
n&$\ 3$&1.5&20&70&\ 60&\ 3.5&$3.0$     &26.8  \cr
o&$\ 1.4$&0.55&20&\ 70&60&10.5&$\ 0.95$&17.3 \cr
\noalign{\vskip2pt\hrule\vskip1pt}
}
\endtable

\beginfigure*3
\line{\psfig{figure=orb1.ps,height=6cm}\hfil
\psfig{file=orb2.ps,height=6cm}\hfil
\psfig{file=orb3.ps,height=6cm}}
\vskip2pt
\line{\psfig{figure=orb1a.ps,height=6cm}\hfil
\psfig{file=orb2a.ps,height=6cm}\hfil
\psfig{file=orb3a.ps,height=6cm}}
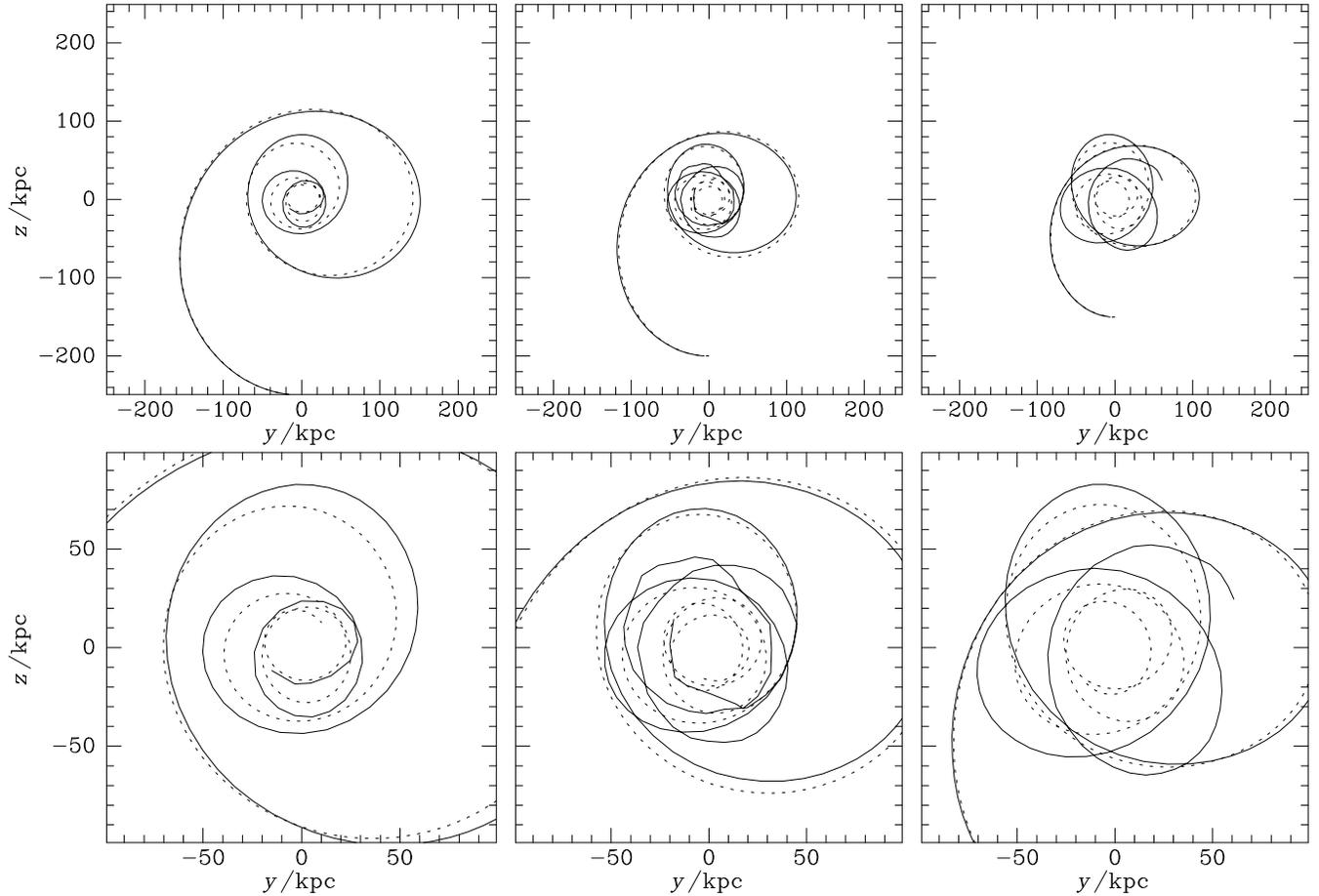
\caption{{Figure 3.} Orbits obtained by full $N$-body
simulation (full curves) and ones obtained with the semi-analytic approximation
(dashed curves). The lower three panels show, on an enlarged scale, the
central  region of the orbit above.}
\endfigure

\beginfigure4
\centerline{\psfig{file=mass1.ps,width=\hsize}}
\vskip2pt
\centerline{\psfig{file=mass2.ps,width=\hsize}}
\vskip2pt
\centerline{\psfig{file=mass3.ps,width=\hsize}}
\caption{{\bf Figure 4.} $M_{\rm D}$ as a
function of $t$ for the orbits shown in Fig.~3 from $N$-body simulations
(full curves) and from the semi-analytic model (dashed curves). From top to
bottom the panels correspond to the panels from left to right in Fig.~3.}
\endfigure

\beginfigure5
\centerline{\psfig{file=sig1.ps,width=\hsize}}
\vskip2pt
\centerline{\psfig{file=sig2.ps,width=\hsize}}
\vskip2pt
\centerline{\psfig{file=sig3.ps,width=\hsize}}
\caption{{\bf Figure 5.} $\sigma_{\rm D0}$ as a
function of $t$ for the orbits shown in Fig.~3 from $N$-body simulations
(full curves) and from the semi-analytic model (dashed curves). From top to
bottom the panels correspond to the panels from left to right in Fig.~3.}
\endfigure

\subsection{Results}

The full curves in Fig.~3 show three possible orbits for the Dwarf.  The
initial configurations from which these simulations start are specified by
columns 2 to 6 and the rows labelled A, F and K of Table 2: column 2 gives
the initial mass that one obtains by integrating the initial density profile
[eq.~\initdprof)] with the values of the characteristic radii given in
columns 4 and 5. Column 6 gives the Dwarf's initial Galactocentric radius.
Columns 3 and 7 to 9 refer to the semi-analytic calculations described
below.  All orbits pass over the Galactic poles and start at apocentre
moving at $103\kms$.  The full curves in Fig.~4 show how the mass of the
Dwarf declines during the simulations. The full curves in Fig.~5 show as a
function of time the one-dimensional velocity dispersion of particles that
are bound to the Dwarf. At late times the curves become jagged because the
number of bound particles becomes small.

\section{Semi-analytic model}

In the semi-analytic model we consider the Dwarf to be a particle of
variable mass that moves in a  fixed potential and suffers drag as
a consequence of dynamical friction.
The equation of motion
to be integrated for the Dwarf is [BT eq.~(7-18)]
$$\eqnam\fricform
{\d\b v\over\d t}=-\nabla\Phi_{\rm G}(r)-4\pi\ln\Lambda
{G^2\rho_{\rm G}M_{\rm
D}\over v^3}\Big[\erf(X)-{2X\over\sqrt{\pi}}\e^{-X^2}\Big]\b v,
\eqno\newe$$
where $\Phi_{\rm G}$ is the Galactic potential and $\sqrt{2}X=v/\sigma$ 
is the ratio of
the Dwarf's speed to the one-dimensional velocity dispersion within the
Galactic halo at the Dwarf's location. For simplicity, we take $\Phi_{\rm
G}$ to be spherically symmetric. The current mass of the Dwarf,
$M_{\rm D}$, is determined as follows. We define
\eqnam\massrt$$
M_{\rm D}(r_\t,t)=4\pi D(t)\int_0^{r_\t}\d r\,r^2\rho_{\rm D}(r,0),
\eqno\firste$$
where $\rho_{\rm D}(r,0)$ is the Dwarf's initial density profile [eq.~\initdprof)]
and 
$$
D(t)={\tanh\big[\beta r_\t(t)/r_\t(0)\big]\over \tanh(\beta)}
\eqno\laste b$$ 
is a ``dilution function'' that takes into account the tendency of tidal
stripping to decrease the density even inside the tidal radius because
some tidally stripped stars initially spent time at small radii. $\beta$ is
a parameter that controls the speed with which the Dwarf is torn apart:
if $\beta$ is large, the stripping of material has little impact on the
inner Dwarf until mass loss from the outside is far advanced.
Equation \refe1a) gives an
estimate of the mass of the Dwarf for any assumed value of $r_\t$. The value
of $r_\t$ is determined at each time-step by finding the radius $r_t'$ that
satisfies the usual tidal
condition
$$
{M_{\rm D}(r_\t')\over r_\t^{\prime3}}={M_{\rm G}(r-r_\t')\over(r-r_\t')^3}
\eqno\newe$$
and then setting $r_\t=\min(r_\t,r_\t')$. Hence, $r_\t$ is not permitted to
increase as the Dwarf moves from peri- to apo-centre. Were it allowed to
increase, equation \refe2a) would give rise to an unphysical increase in
$M_{\rm D}$ because it neglects the sharp fall in the density that will in
reality be encountered outside the smallest value that  $r_\t$ has taken
along the orbit: mass lost is lost for ever.

The full curves in Fig.~3 show three orbits for the Dwarf that were followed
by full $N$-body simulation. The dashed curves show orbits integrated from
the same initial configurations using the semi-analytical model with
$\ln\Lambda=8.5$ and $\beta=3.7$ in equation \refe2b). Although the initial
conditions corresponding to the panels differ considerably, the
semi-analytic model provides a good qualitative fit to all three until the
dissolution of the Dwarf is far advanced. The fit is best for the left-hand
orbit, in which a very massive Dwarf starts from a large apocentre. This is
the case in which the $N$-body calculation should be most reliable because
at any given time there are always more particles in the Dwarf in this
simulation than in either of the others. The major shortcoming of the
semi-analytic model is that it makes the Dwarf's orbit unrealistically
circular at late times (see the lower panels of Fig.~3). This defect arises
because Chandrasekhar's dynamical-friction formula \fricform) makes the drag
proportional to the local density, whereas the response to a satellite is in
reality global (Hernquist \& Weinberg, 1989). In consequence, equation
\fricform) over-emphasizes dragging at pericentre, where energy is
relatively more important than angular-momentum loss.

Fig.~4 shows, for the orbits shown in Fig.~3, plots of $M_{\rm D}$ as a
function of $t$. The full curves are obtained from the $N$-body simulations,
and the dashed curves from the semi-analytic model. Again we see that the
semi-analytic model reproduces the $N$-body data well, especially for the
most massive Dwarf. The main difference between the semi-analytic and
$N$-body curves is a tendency for $M_{\rm D}$ to be constant after each
pericentre in the semi-analytic case. Towards the end of each simulation,
when there are only $\sim100$ particles in the Dwarf, two-body relaxation
will artificially hasten the demise of the Dwarf.  Hence the tendency of the
semi-analytic models to have larger masses at late times than the $N$-body
models is not necessarily a defect of the semi-analytic models.

The dashed curves in Fig.~5 show the velocity dispersion at $r=1\kpc$,
$\sigma_0$, in the semi-analytic models. This was calculated by integrating
the Jeans equation for an isotropic spherical system [BT eq.~(4-54)] with
the density profile set equal to $D\rho_{\rm D}(r,0)$ for $r<r_\t$ and zero
otherwise.  Comparison of Figs 4 and 5 shows that this prescription
over-estimates the velocity dispersion in the corresponding $N$-body model
when the semi-analytic model over-estimates the model's mass, and vice-versa
when the mass is underestimated. Therefore, the velocity-dispersion values
returned by the semi-analytic model probably provide fairly reliable guides
to the velocity dispersion of the Dwarf, especially at late times, when
small-number statistics make the velocity dispersions of the $N$-body models
very uncertain.

\subsection{Results}

Table 2 summarizes a series of orbits computed using the semi-analytic
model. The initial conditions are chosen to illustrate the constraints that
must be satisfied if the Dwarf is first to reach a galactocentric radius
$r=16\kpc$ in a time, $t_{\rm sink}$, of order $11\Gyr$. The column headed
$M_{\rm D\infty}$ is the mass one obtains by integrating the density profile
\initdprof). The actual initial mass, $M_{\rm D}(0)$ given in the next
column, is usually somewhat smaller because equations \massrt) imply that
$r_\t<r_\infty$ even at the start of the computation. The columns headed
$r_{\rm D0}$ and $r_{\rm D\infty}$ give the values for the dwarf of the
characteristic radii $r_0,r_\infty$ of equation \initdprof). $R_{\rm D}(0)$
is the Dwarf's initial Galactocentric radius, $t_{\rm sink}$ is the time at
which it first reaches $16\kpc$ from the Galactic centre, and $M_{\rm
D}(t_{\rm sink})$ is its mass at that instant. The final column gives the
corresponding velocity dispersion $1\kpc$ from the Dwarf's centre calculated
from the Jeans equation and the density profile $D(t_{\rm sink})\rho(r,0)$
for $r<r_\t$ as described above.

At the start of the first two orbits, the Dwarf contains $10^{11}\msun$ and
starts from Galactocentric radius $r=250\kpc$. The orbits differ in the
distribution of mass within the Dwarf: at the start of orbit (b) the Dwarf
is more extended than at the start of orbit (a). Consequently, the Dwarf is
more rapidly stripped on orbit (b), suffers less dynamical friction, and
takes longer to reach the inner Galaxy. At the end of each orbit the Dwarf
has been stripped down to $\lta2\times10^9\msun$, a mere 2 percent of its
initial mass. When $300\,000$ particles are used to represent the Galactic
halo, this mass  corresponds to $\lta240$ particles in the Dwarf.

Orbit (c) starts from the same configuration as orbit (a), but at a smaller
initial radius, $r=225\kpc$. Consequently, it reaches $r=16\kpc$ sooner:
after $t_{\rm sink}=9\Gyr$. If the Dwarf is more extended, $r=16\kpc$
is reached at a later time, as on orbit (d). Orbits (e) and (f) show
a similar progression as one moves in to an initial Galactocentric radius
$r=200\kpc$.

Orbit (g) shows that the time required to reach $16\kpc$ from $200\kpc$ can
be decreased by making the Dwarf more compact, even though of a lower mass.
Orbit (h) shows the effect of reducing the mass while holding constant the
shape of the density profile: a 27 percent decrease in the mass increases
$t_{\rm sink}$ by 35 percent. Orbit (i) reaches $r=16\kpc$ from $r=200\kpc$
in the target time. Orbits (j) to (o) indicate the initial mass and shape
required to reach $r=16\kpc$ from $r=150$ to $60\kpc$ in the target time.

A constraint on possible orbits is provided by the measured velocity
dispersion of the Dwarf's stars: $\sigma_0=(11\pm 0.7)\kms$ (Ibata et al.\ 1997).
This may be compared with the velocity dispersions calculated from the
semi-analytic model as described at the end of the previous subsection,
which are listed in the last column of Table 2.  It is striking that
although the initial mass declines by a factor $\sim20$ from top to bottom
through the table, the final velocity dispersion lies in the range $19.4$ to
$24\kms$ for models in which $t_{\rm sink}=10\pm1\Gyr$. Two effects must
contribute to the significant difference between this number and the smaller
observational value. Most obviously, the model value refers to all the
Dwarf's material, which is predominantly dark. Since the dark matter is more
extended, its central velocity dispersion must be larger than the value
measured for the luminous matter. A second effect is velocity anisotropy:
the model assumes velocity isotropy, which almost certainly does not hold.
The measured value is for a line-of-sight almost perpendicular to the
Dwarf's long axis, which by the tensor virial theorem should be lower than
the line-of-sight value for the corresponding isotropic model. If both of
these effects caused the observational value to be smaller than the
calculated value by a factor $\sim1.2$, the difference between theory and
observation would be fully accounted for. It is plausible that the factors
are of this order, though only substantially more sophisticated modelling
could reliably determine them.

\beginfigure6
\centerline{\psfig{file=sgrorb.ps,width=\hsize}}
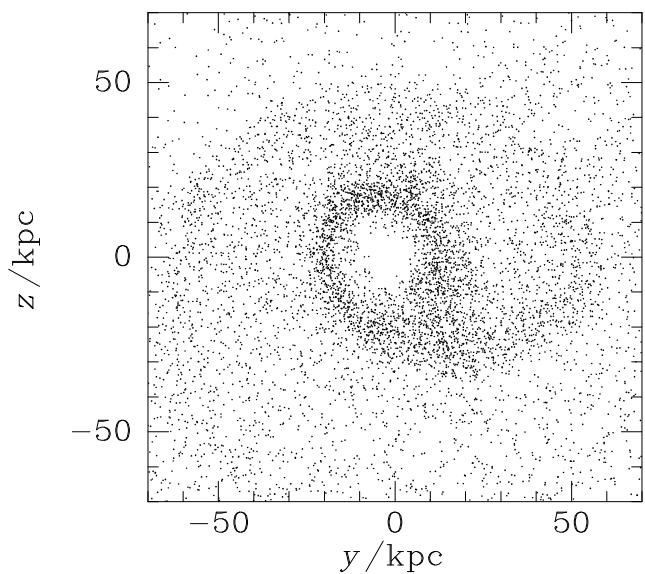
\caption{{\bf Figure 6.} The distribution of Dwarf particles projected onto
the plane perpendicular to the Dwarf's orbit.}
\endfigure

\section{Conclusions}

We have modelled the orbital decay of the Sgr Dwarf under the assumption
that the Galactic halo extends  to $250\kpc$. A mixture
of $N$-body simulations and semi-analytic modelling suggests that the
present configuration of the Dwarf is consistent with a wide variety of
orbital histories. At one extreme the Dwarf starts out with a mass
$\sim10^{11}\msun$ at a Galactocentric radius $\gta200\kpc$. At the other
extreme it starts out with a mass $\sim1.2\times10^9\msun$ at a radius
$\sim60\kpc$, as suggested by Ibata \& Lewis (1998). The larger the initial
Galactocentric distance of the Dwarf, the more massive the Dwarf must have
been initially. The final velocity dispersion of the Dwarf is remarkably
insensitive to the Dwarf's history, because as the initial distance of the
Dwarf is increased, its dark halo has to be extended at an approximately
constant velocity dispersion, in order to maintain an appropriate rate of
orbital decay and tidal stripping by the Galactic potential, which has
nearly constant velocity dispersion by construction.

If the Galaxy has stripped $\gta10^{10}\msun$ from the Dwarf, the stripped
material should form a complete ring around the sky -- see Fig.~6.
Unfortunately, the great majority of the mass plotted is dark and not
directly observable. Luminous matter was originally confined to the centre
of the Dwarf and will have been stripped in quantity only recently. Hence it
will be less uniformly spread over the sky than the dark matter. 

In general dark matter can only be detected through its gravitational field.
Fortunately, we have a sensitive gravitational field detector in place: the
outer Galactic disk. We hope soon to report on how the Galaxy's outer HI
disk responds to the time-dependent gravitational field that is generated by
both the Dwarf and the Magellanic Clouds.  It is already apparent that the
more massive Dwarf models generate distortions of the disk that are
comparable in magnitude to the observed Galactic warp. What is still unclear
is whether the Dwarf and Clouds can between them explain the particular
morphology of the Galactic warp.


\section*{References}

\beginrefs
\bibitem Binney J.J., Tremaine S.D., 1987, `Galactic Dynamics', Princeton
University Press, Princeton (BT)

\bibitem Bontekoe T.R., 1988, Ph.D. thesis Groningen University

\bibitem Dehnen W., \& Binney J.J., 1997, \mn 294 429

\bibitem G\'omez-Flechoso M.A., Fux R., Martinet L., 1999, astro-ph/9904263


\bibitem Hernquist L., Weinberg M., 1989, \mn 238 407

\bibitem Ibata R.A., Gilmore G., Irwin M.J., 1994, Nature 370, 194

\bibitem Ibata R.A., Lewis G.F., 1998, \apj 500 57

\bibitem Ibata R.A., Wyse F.G., Gilmore G., Irwin M.J., Suntzeff M.B., 1997,
	\aj 113 634

\bibitem Jiang I.-G., Binney J.J.,  1999, \mn 303 L7

\bibitem Johnston K.V., Spergel D.N., Hernquist L., 1995, \apj 451 598



\bibitem Quinn T.R., Katz N., Stadel J., Lake G., 1997, astro-ph/9710043

\bibitem Schmoldt I.M., Saha P., 1998, \aj 115 2231

\bibitem Vel\'asquez H., White S.D.M., 1995, \mn 275 L23

\bibitem Zhao H.-S., 1998, \apj 500 L149

\endrefs
\bye